\documentstyle[11pt,epsfig]{article}

\newlength{\dinwidth}
\newlength{\dinmargin}
\def\lapproxeq{\lower .7ex\hbox{$\;\stackrel{\textstyle
<}{\sim}\;$}}
\def\gapproxeq{\lower .7ex\hbox{$\;\stackrel{\textstyle
>}{\sim}\;$}}

\def\be{\begin{equation}}
\def\ee{\end{equation}}
\def\bea{\begin{eqnarray}}
\def\eea{\end{eqnarray}}

\def\lesim{ \;\raisebox{-.7ex}{$\stackrel{\textstyle
<}{\sim}$}\; }

\def\lapproxeq{\lower .7ex\hbox{$\;\stackrel{\textstyle
<}{\sim}\;$}}
\def\gapproxeq{\lower .7ex\hbox{$\;\stackrel{\textstyle
>}{\sim}\;$}}
\def\be{\begin{equation}}
\def\ee{\end{equation}}
\def\bea{\begin{eqnarray}}
\def\eea{\end{eqnarray}}
\def\bea*{\begin{eqnarray*}}
\def\eea*{\end{eqnarray*}}

\def\GeV{{\rm GeV}}

\begin{document}
\begin{flushright}
\end{flushright}

\begin{center}
{\Large \bf {\bf The Visibility of Shell-Type Supernova Remnants in Gamma Rays}}

\vspace*{0.5cm}
\textsc{A.D. Erlykin$^{a,b}$ and A.W. Wolfendale$^b$} \\

\vspace*{0.5cm} $^a$ Lebedev Physical Institute, Leninsky
Prospekt, Moscow, Russia \\
$^b$ Department of Physics, University of
Durham, DH1 3LE, UK \\
\end{center}

\begin{abstract} The question of the origin of the cosmic radiation
(CR) is a continuing one. The idea that the shocks from supernova
remnants (SNR) expanding into the interstellar medium (ISM) accelerate
CR is still a
popular one but a number of authors have drawn attention to the
fact that the experimental evidence for the presence of gamma rays
from the expected interaction of CR with gas in the remnant is poor.
Indeed, it is claimed that many SNR are not `seen' in GeV or TeV
gamma rays, whereas `they should have been'.

We have looked at this problem and we conclude that the idea of CR
production in SNR cannot be faulted in this way, if the evacuation of
ambient gas by the stellar wind of the progenitor star and,
frequently, by associated earlier close-by SN, is taken into account; such 
phenomena are expected for the important Type II SN which result from 
very massive stars and which provide the SNR which are thought to 
accelerate CR.

Other SNR have, apparently, been seen and the interaction of
SNR-accelerated particles with adjacent molecular clouds has been
deemed responsible. However, we worry about this interpretation
because of the slow progress of the SNR shock through such clouds,
although electron-effects may, indeed, contribute.

This paper is devoted mainly to the visibility of nearby
(~within about 1 kpc~) SNR in gamma rays although many of the
arguments also relate to remote SNR. For the nearby SNR another problem enters
the scene: the large angular spread of the remnant. It is
especially important for the old SNR, where cosmic rays have
already diffused to a large distance from the SNR center. We have also
examined the effect of the 'anomalous diffusion' of CR propagation in the
non-homogeneous interstellar medium on the visibility of SNR
for gamma rays of different energies.
\end{abstract}

\section{Introduction} \label{sec:intro}

The idea that CR of energy up to the `knee' in the spectrum (at $\sim
3$~PeV) are
produced by SNR shocks is well known and there is considerable
literature on the subject. We ourselves have taken these models
and examined a number of CR `results' (anisotropy, mass
composition, spectral variations, etc.) and found a satisfactory
outcome. However, others (e.g.~Plaga (2001), Pohl (2001), Torres
et~al. (2002), V\"{o}lk (2002)~) have drawn attention to the lack of a 
gamma ray
signal from many known SNR, the gamma rays being expected to
arise from CR interactions with the gas (~and the radiation field~) 
in the SNR. The interactions are bremsstrahlung and inverse compton 
scattering (~IC~) for electrons and pion production
for protons and heavier nuclei. Clearly this is an unsatisfactory
situation and one needing an in-depth examination.

The question of the relative contributions to the gamma ray flux from
protons (~nuclei~) via $\pi^\circ$-decay and electrons is one that
pervades the subject. At `low' energies (~eg. $E_\gamma >$ 0.1 GeV~)
it is generally regarded that in the diffuse gamma ray flux, at least,
only some 20\% comes from electrons (~see Ramana Murthy and
Wolfendale, 1993 for a summary of the estimates of their quantity~).

In the TeV region the answer is less clear cut insofar as many workers
have argued that the IC contribution may surpass that from
$\pi^\circ$-decay. Two points are relevant here: \\
(i) In one sense, the IC contribution is not important if, as is the
case here, our goal is to examine the argument that `lack of observation
of TeV gamma rays shows that CR in general are not produced by the
objects in question: SNR'. This is the argument that is adopted
here. \\
(ii) In fact, there are some who argue that for some sources at least,
IC from electrons is not important. Berezhko et al. (~2001, 2003~) adopt
this position; for their models of SNR 1006 and CasA, they find that
IC provides
only $\sim$30\% and 1\%, respectively of the total flux of gamma rays at 1 TeV.

These two points allow us to ignore the IC contribution, although we
mention its effect further when considering the TeV gamma ray results
for the SN just mentioned in \S 5.3.1.

Gamma-ray production in shell-type SNR has been widely discussed and an extended list of references can be found
in Torres et al. (2002). The form of our examination here is, in
part, as follows:
\begin{itemize}
\item[(i)] We derive the expected proton intensity vs. distance from
the SNR and the gamma ray intensity vs.
angle from an SNR for two energy thresholds: 0.1~GeV and 1~TeV. The
derivations are for `standard' conditions of ISM density, shock
strength, etc., and they relate to different distances and to
different ages of SNR.
\item[(ii)] The experimental situation is briefly examined, from the
standpoint of the searches made and the fluxes, and upper limits,
recorded.
\item[(iii)] A detailed study is made of the likely conditions in
the actual ISM, not least the density of the target ISM and thereby 
the predicted fluxes for the two
energy thresholds and the various distances and ages. The effects of
the non-homogeneous structure of the ISM and the presence of molecular
clouds are also studied.
\item[(iv)] A comparison of `observed and expected' fluxes is
given.
\item[(v)] The conclusion draws the strands together.

\end{itemize}

In another paper we examine the situation for gamma rays from our proposed
'single source' (~see, for example, Erlykin and Wolfendale, 1997~).

It will be apparent that the present work includes new features, not
previously considered, at least in the manner whereby we endeavour to
use actual Galactic properties - most notably those of the
gas in the ISM and the way in which particles propagate.

\section{The expected gamma ray signal from SNR, for `standard
conditions'} \label{sec:expectedsignal}

Our intention is not to examine the precise theoretical model of the
SN explosion and development of the SNR and the consequent particle 
acceleration, which has been the subject of much work by others, but
rather to draw a simple scenario, which nevertheless satisfies the
basic physical principles and energy requirements. Where there {\em is} a
new analysis is in the case of diffusion of the particles after
leaving the remnant, where `anomalous diffusion' is considered as a variant.

\subsection{The acceleration model} \label{sec:accelerationmodel}

We follow our model enunciated in Erlykin and Wolfendale (~referred to
henceforth as EW~)
(2001). In it we use the treatment of Axford (1981),
complemented by the numerical calculations of Berezhko (1999), in
which the SNR shock propagates through the tenuous hot ISM with
density $n \approx 3\cdot 10^{-3} {\rm cm}^{-3}$ (~a more recent
estimate is $4\cdot 10^{-3} {\rm cm}^{-3}$, Ferri\`{e}re, 2001~). We have
chosen these conditions because we are interested in the nearby
SNR and they correspond to the conditions in our local
superbubble. In fact, the effect of the ISM density in the calculation
of the proton distribution is not too important insofar as we
standardise the spectra to correspond to a fixed total CR energy. 
In what follows we give some details about the model only insofar as
some features of the predicted gamma ray yield are sensitive to it.

With the density indicated the radius of the shock wave,
$R_s$, is given by
\be \label{eq:R_s} R_s=R_0\sqrt{t/T_0}, \ee
where $R_0=50$~pc and $T_0=2\times10^4$~y.

In fact, the exponent (~0.5~) differs from the conventional value,
0.4, for the Sedov region; one reason has been given by Berezhko,
(~1999~) and is related to the effect of cosmic ray
pressure. Interestingly, Moffett et al., (~1993~) quote 0.48 from
direct measurements for SN 1006. In any event, the exact value is not
very important for the reason given above, viz. energy normalization 
is applied. 

A similar remark can be made about the fact that $R_0$ depends on
density whereas in what follows we disregard this dependence. The
logic is in part the energy argument just referred to and in part the
fact that the operative parameter is the time spent in the remnant
before the `bubble bursts', i.e. the energy density falls to roughly
the ambient value. There are compensating factors which make this time
only weakly dependent on gas density, at least for the range of
densities considered here. 

Particles are
accelerated such that the CR energy density is related to the time
from the SN explosion by
\be \label{eq:rho_CR} \rho_{\rm E}\propto M^2(t), \ee
$M(t)$ being the Mach number, which, following (\ref{eq:R_s}), decreases
with time as
\be \label{eq:M(t)} M(t)\propto t^{-\frac{1}{2}}. \ee
At each moment the cosmic rays are produced with a power-law
rigidity spectrum in the rigidity interval $0.1$--$4 \times
10^5$~GV (Berezhko et~al., 1996). The spectral exponent at time
$t$ is given by
\be \label{eq:gamma(t)} \gamma(t) = \frac{2+2/M^2(t)}{1-1/M^2(t)} \ee

It is assumed that CR take, finally, $10^{50}$~erg, i.e. 10\% of the
available kinetic energy. The remnant is
assumed to `release' the particles at a radius of $R_s=100$~pc at
a time $t=8\times10^4$~y. In the initial calculations the
particles were assumed to be protons in this stylised model,
although later we included other nuclei. It can be added that adiabatic
losses are `allowed for' by taking a specific value for the final CR
energy on emergence from the remnant.

\subsection{Particle propagation within and beyond the remnant}
\label{sec:particlepropagation}

\subsubsection{Within the remnant} \label{sec:within}

It is assumed that the particles diffuse rapidly after
acceleration so that they fill the remnant uniformly and the
intensity is thus independent of radius at any instant. The
argument is that, despite the mean free path for scattering at GeV
energies being very short---and thus one might expect a
concentration of such particles near the centre of the
remnant---turbulence will cause mixing.

\subsubsection{Outside the remnant} \label{sec:outside}

The particles emerging from the remnant will diffuse away and
the majority will eventually escape from the Galaxy. Their intensity 
versus distance is an important datum for two reasons:
\begin{itemize}
\item[(i)] There will be a `halo' of gamma rays round the SNR, and
some, at least, of these gamma rays will often count as having come
from the SNR.
\item[(ii)] In the analysis of gamma rays expected from our Single
Source (~the nearby SNR~)
it is those particles which have diffused from this SNR, that give the 
`peak', which causes the knee in the primary energy spectrum. This is
the point that Bhadra (~2002~) has addressed in his analysis of the
extent to which the gamma ray signal from our Single Source might be
expected to have been observed.
\end{itemize}

Calculations have been made for two modes of propagation of the CR
particles: 
the `normal', gaussian form and that for `anomalous diffusion'. In the
latter analysis we follow the work of Lagutin et~al. (2001a,b),
Erlykin et~al. (2003) and EW (2002a)). Arguments in favour of anomalous
diffusion have been advanced by Erlykin et~al. (2003) and are mainly the
following:
\begin{itemize}
\item[(i)] the spatial distribution of matter, magnetic and radiation
fields in the ISM is highly irregular and non-homogeneous;
\item[(ii)] in many cases the spectra of the irregularities have a power-law
character, which favours the fractal structure of the ISM;
\item[(iii)] anomalous diffusion in the fractal ISM allows an
understanding of
a number of observed effects, including: the so-called `GeV-excess' of gamma rays in
the Inner Galaxy, the small radial gradient of low energy cosmic rays,
the observation of a Galactic Plane Enhancement in the Outer Galaxy at
energies about $10^5$ GeV and the softer energy spectrum of cosmic rays
in the galactic halo.
\end{itemize}
Using the nomenclature of Lagutin at al.,(~2001a,b~) the mode of
propagation is characterised by a parameter
`$\alpha$', where $\alpha = 2$ for normal `gaussian' diffusion and
takes smaller values for anomalous diffusion. We have made the case in EW
(2002a) for $\alpha=1$ locally. The
extent to which the remnant itself affects the mode of propagation
just outside the remnant will be considered later.

The difference between the two modes appears in the shape of the
lateral distribution function for the cosmic ray intensity:
$\frac{1}{(1+x^2)^2}$ for $\alpha = 1$ and $\exp(-\frac{x^2}{4})$
for $\alpha = 2$, with $x = \frac{r}{R_d}$, $r$ being the
distance from the radius $R_s = 100$ pc where the particles start
to diffuse and $R_d$ being the diffusion radius which is defined as
$R_d = H_z(\frac{t}{\tau(E)})^{\frac{1}{\alpha}}$, i.e., there is 
a different time dependence for the two modes. $H_z = 1$~kpc 
(~as assumed previously~) for
the vertical scale of the galactic halo, and $t$ and $\tau(E)$ are the
diffusion time and proton lifetime against escape, respectively.
Clearly, the distribution function for the anomalous diffusion case
falls off much more slowly than for the gaussian, at large {\em x}.

In the calculations we again follow EW (2001) and adopt
\be \label{eq:tau} \tau = 4\times10^7\ E^{-\delta}\ {\rm y} \ee
Here, and below, $E$ is in GeV.

Since the scattering on magnetic irregularities has a predominantly
resonant character, i.e., it is most efficient for particles
with giroradius equal to the size of the irregularity 
(~Berezinsky et~al. 1984, Longair, 1992~), then the exponent
$\delta$ is connected with $\alpha$ as
\be \delta = \alpha/2 \ee
(Erlykin et~al., 2003). We have, however, adopted $\delta = 0.5$
for both modes to reveal the difference between them due to the
first two factors: the diffusion front shape and the time
dependence of the diffusion radius.

We illustrate the difference between the two modes of propagation by
two Figures. Figure~1  shows the results for the time dependence
of the cosmic ray energy density at different distances from the
SNR center.
\begin{figure}[hptb]
\vspace{0.5cm}
\begin{center}
\includegraphics[height=12cm,width=15cm]{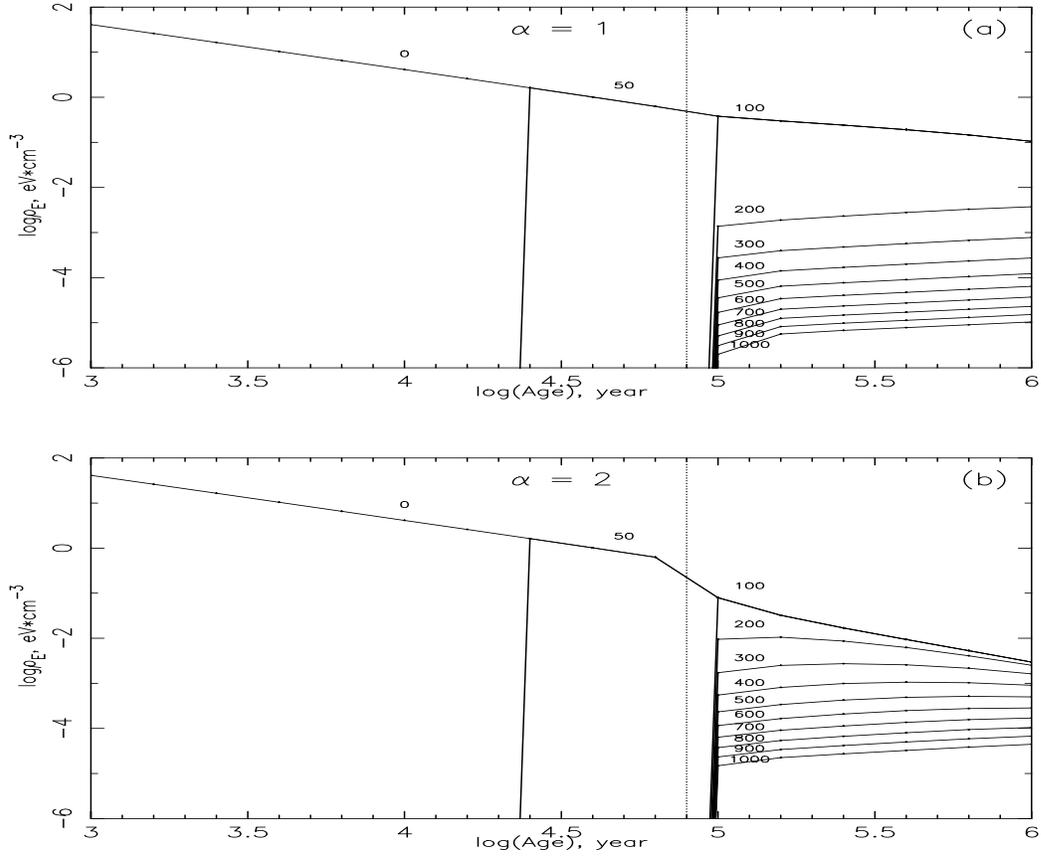}
\vspace{1.0cm} \caption{\footnotesize Energy density of cosmic
rays from SNR of different ages, at different distances from their
center and for `anomalous'~(a) and `normal'~(b) diffusion. Numbers
on the curves indicate the distance from the SNR center in pc. The
vertical dotted line at $8\cdot 10^4$ years marks the time of the end
of the expansion phase. The sudden changes in the energy density are
due to the `bin widths' used in the calculations.}
\label{fig:vis1}
\end{center}
\end{figure}
The higher energy density for the `normal' diffusion at distances
larger than 100~pc from the SNR centre compared with the `anomalous'
one is due to the faster diffusion for $\alpha = 2$ at relatively
small times after it starts.

Figure~2 shows the lateral distribution functions for the two modes of
propagation and for two proton energies. The energies considered are
those roughly relevant to the gamma ray energy thresholds: 2~GeV
for $E_\gamma>0.1\ \GeV$ and 10~TeV for $E_\gamma>1$~TeV. The
difference between the cases of $\alpha=1$ and $\alpha=2$ is seen
to be quite marked.
\begin{figure}[hptb]
\includegraphics[height=15cm,width=15cm]{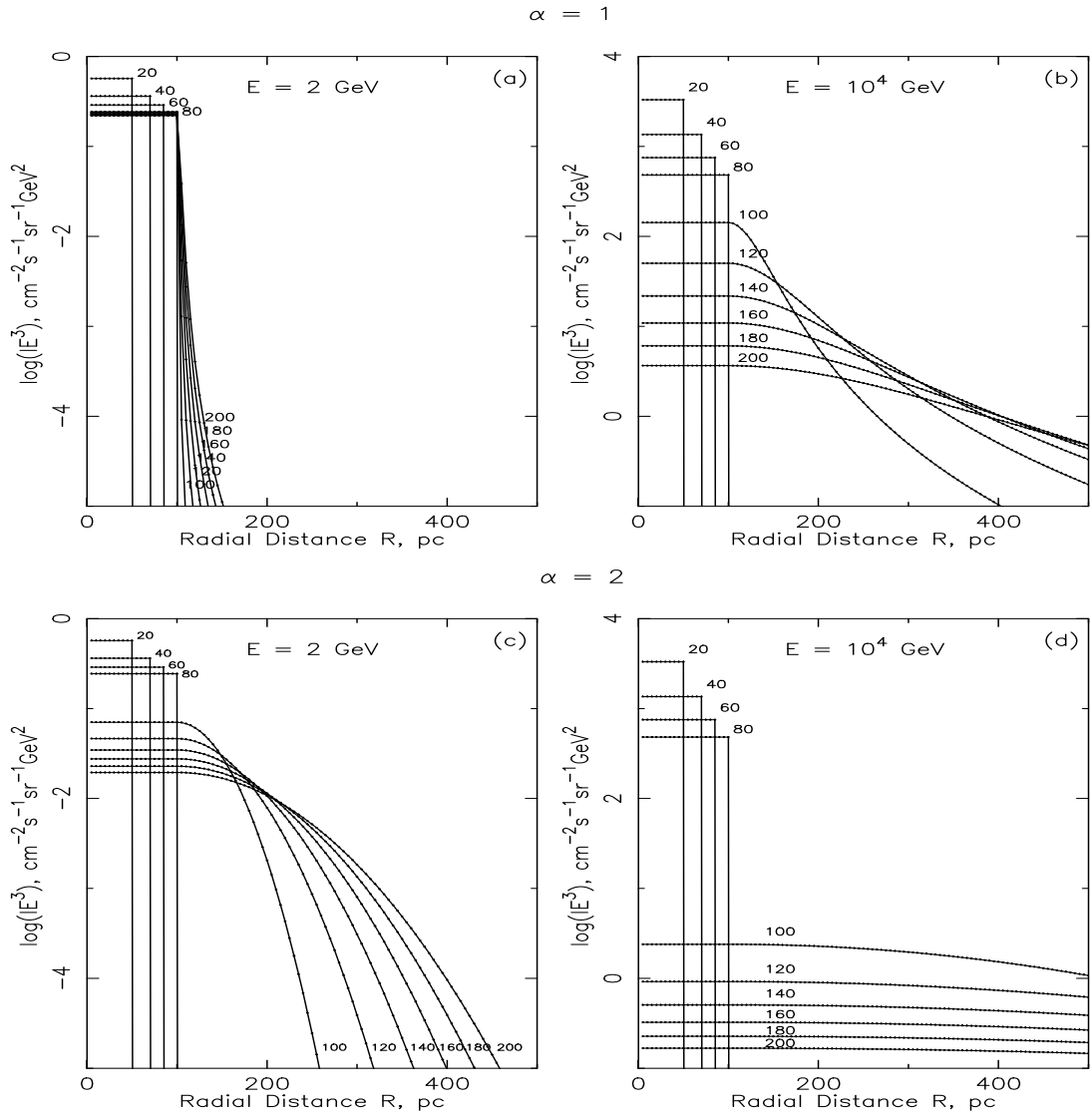}
\vspace{1cm} \caption{\footnotesize Lateral distribution function
(LDF) of cosmic rays from SNR of different ages. The numbers on the curves
show the SNR age in kyears. Figures (a) and (b) show the LDF for
$\alpha = 1$, (c) and (d) for $\alpha = 2$, (a) and (c) for 
energy $E = 2$~GeV, (b) and (d) for $E = 10^4$~GeV.}
\label{fig:vis2}
\end{figure}

\subsection{Gamma Ray Emissivity} \label{sec:emissivity}

\subsubsection{\bf $pp$-collisions } \label{sec:emisp}

The emissivity, in terms of number of gamma rays produced above a
particular energy, per proton--ISM proton ($pp$) interaction, and
its conversion to emissivity for the ambient CR spectrum, has been
given by many authors, following the early work of Stecker (1971).

We derived these values on the basis of the following assumptions:
\begin{itemize}
\item[(i)] the gamma quanta come from the decay of $\pi^\circ$'s;
\item[(ii)] the inclusive spectrum of $\pi^\circ$'s in the C-system of
$pp$-collisions has the gaussian form $\frac{dn}{dy} =
A\exp\left(-\frac{y^2}{y_0^2(k_{\gamma})}\right)$, where $y$ is
the rapidity;
\item[(iii)] the partial inelasticity $k_\gamma$ does not depend on
the primary energy and is equal to 0.15;
\item[(iv)] the total
multiplicity of gamma quanta and its density in the central
rapidity region is close to those observed in experiments.
\end{itemize}

\begin{figure}[hptb]
\begin{center}
\includegraphics[height=15cm,width=15cm]{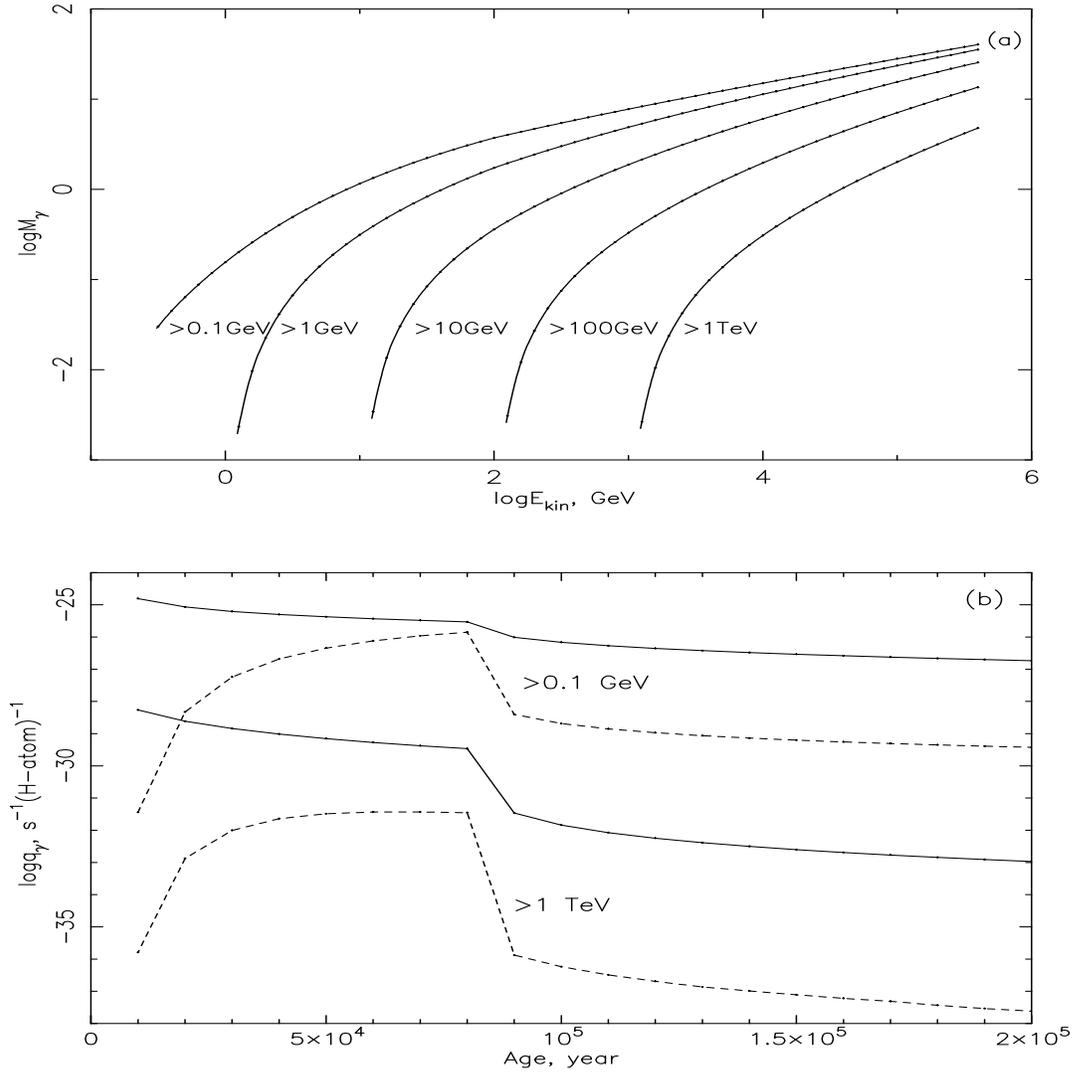}
\vspace{1cm} 
\caption{\footnotesize (a) The multiplicity of 
gamma quanta produced in $pp$-collisions for different threshold
energies, indicated by numbers near the curves, as a function of
the primary proton energy. (b) The gamma ray emissivity for cosmic
rays in the SNR center for two threshold energies: 0.1~GeV and
1~TeV as a function of SNR age: full lines - for $pp$-collisions,
dashed lines - for electron bremsstrahlung.} 
\label{fig:vis3}
\end{center}
\end{figure}

The gamma ray multiplicity $M_\gamma$ as a function of the primary proton
energy in the energy range $0.1$--$4\times10^5$~GeV, where
protons are accelerated after the SN explosion, is shown in Figure 
\ref{fig:vis3}a
for different energy thresholds $E_\gamma^{\rm thr}$ of gamma
quanta. For $E_\gamma^{\rm thr} = 0.1$~GeV the results agree with
the total multiplicity and particle density in the central region
of rapidity measured in the accelerator UA5 experiment (Alpgard
et~al., 1982); for $E_\gamma^{\rm thr} = 1$ TeV they agree with the results
of the UA7 experiment (Pare et al., 1990).

The gamma ray emissivity depends on the cosmic ray energy spectrum
$I(E)$, the inelastic cross-section for particle collision
$\sigma(E)$, and the gamma ray multiplicity $M_\gamma(E,>E_\gamma^{\rm
thr})$, as
\be q(>E_\gamma^{\rm thr}) = 4 \pi \int
I(E)\sigma(E)M_\gamma(E,>E_\gamma^{\rm thr})dE
\label{eq:emisp} \ee
The energy spectrum depends in
turn on the location and the age of the SNR. As an illustration we
show in Figure \ref{fig:vis3}b the time variation of the emissivity for cosmic
rays in the center of the SNR. At the end of the expansion at the age
of $0.8\times10^5$ years, when the spectrum inside the SNR has a
slope of about 2.15, our values if referred to the energy density
contained in the spectrum agree well with the results of Drury
et~al.,(1994); these results are frequently adopted.

\subsubsection{\bf $Ap$-collisions} \label{sec:emisA}

For $Ap$-collisions the cross section $\sigma_{Ap}$ in
(\ref{eq:emisp}) is larger than the corresponding cross-section
$\sigma_{pp}$ for $pp$-collisions. At the same energy per nucleon
$E_n$ the multiplicity of gamma quanta produced by one `wounded',
i.e. non-spectator nucleon, is the same as for protons. Hence the
total number of gamma quanta produced in a single $Ap$-collision,
$M_\gamma^{Ap}$, should depend on the number of `wounded' nucleons
$n_w$ as $M_\gamma^{Ap} = n_w M_\gamma^{pp}$.

Since we assume that particles of any charge $Z$ and mass $A$ are
accelerated in the same interval of rigidity then if the
corresponding spectrum of nuclei in terms of the energy per
nucleon has a power law character with differential exponent
$\gamma$, it is
\be I_A(E_n) = \left(\frac{Z}{A}\right)^{\gamma - 1}I_p(E_n)
\label{eq:spec} \ee
In the Glauber geometric approach to the $Ap$-collision,
$n_w = A \frac{\sigma_{pp}}{\sigma_{Ap}}$, therefore
\be q_A=4\pi \int \left(\frac{Z}{A}\right)^{\gamma-1}I_p(E)
\sigma_{Ap} A \frac{\sigma_{pp}}{\sigma_{Ap}} M_\gamma^{pp}
dE = \left(\frac{Z}{A}\right)^{\gamma-1} A q_p
\label{eq:emisA} \ee

In the important case of cosmic rays emerging at the end of the
SNR expansion with $\gamma \approx 2$ the gamma ray emissivity
of nuclei is about Z times higher than for
protons {\em with the same rigidity}. Of course, the CR energy input
will be higher for the case where nuclei are concerned.

\subsubsection{Inclusion of the other nuclei in CR and the ISM}

If it is assumed that the mass
composition of cosmic rays produced by SN is similar to that observed 
near the Earth (~Ito, (1988), Wiebel-Sooth and Biermann, (1999)), then
the increase of gamma ray emissivity due to the mixed CR mass composition
is about 1.64$\pm$0.03 for sub-GeV energies and 1.97$\pm$0.05 at 
$\sim$10 TeV energy.

In addition to this factor there is also the effect of non-hydrogen 
nuclei in the
ISM. Dodds et al.,(1976) give, as the multiplying factor 1.40. Thus,
the increase for all CR-ISM over CR {\em p}-H nuclei interactions is 
2.30$\pm$0.04 at sub-GeV energies and 2.76$\pm$0.07 at 10 TeV.

\subsubsection{Contribution of electrons} \label{sec:emise}

Irrespective of arguments about the magnitude of the IC contribution
(~\S1~) - and despite the remarks there about the small magnitude of
the $e$-bremsstrahlung contribution - we have checked the latter.

A finite electron contribution is expected because shock waves
propagating through the hot ISM and plasma accelerate not only protons 
and nuclei, but also electrons, and we have therefore 
evaluated the contribution of electrons to the emissivity of gamma
quanta. At this stage we took into account only bremsstrahlung
because we are interested in the matter effects and the intensity of this
process is proportional to the matter density, likewise the intensity
of $\pi^\circ$ - production. The contribution of synchrotron radiation and
inverse Compton scattering requires the introduction of two additional
input parameters: the strength of the magnetic and radiation
fields. They were accounted for in the total energy losses including
ionization using just the parameters typical for the standard
ISM: $B = 3 \mu G$ and $w_{em} = 0.95 {\rm eV}{\rm cm}^{-3}$.
The former may be too low in a young remnant.

For the evaluation of electron bremsstrahlung we used the same
program, which was developed in EW (2002b), and which let us
explain the low $e/p$ ratio and the steeper energy spectrum of
electrons, compared with protons, as the consequence of the
inefficiency of electron injection for high Mach numbers in SNR
shocks. For the calculation of bremsstrahlung we used the classic
non-screening cross-section given by Rossi, (1952).

The bremsstrahlung photon emissivity is shown in Figure 3b by dashed
lines. The difference in the emissivity between electrons and protons
at the initial stages of the SNR development is due to our assumption
about the inefficiency of high Mach numbers for electrons, which was
absent for protons. It is seen that nearly all the time the
contribution of electrons to the gamma-ray emission is negligible
compared with protons, so that in what follows we ignore the electron 
contribution. It should be remarked, however, that there may well be
regions where enhanced electron acceleration occurs, perhaps due to
non-linear wave generation effects and turbulence in addition to the
degree of turbulence adopted here. Such behaviour may be found in
highly shocked region of molecular clouds (~eg Bykov et
al., 2000~).

\section{Gamma Ray Fluxes, Expected and Measured} \label{sec:vsangle}

\subsection{Expected gamma ray fluxes}
\label{sec:expected}

In Figure 4 we show the total gamma ray flux from a nearby SNR as a
function of its age (~for primary CR protons~). The datum matter density
everywhere has been taken as 1 H-atom~cm$^{-3}$ (~heavier nuclei in
the CR beam and in the ISM are not included~). 
At the maximum of the emission the fluxes can be approximated as
\be
F_\gamma^{max}(>0.1 GeV) = 3.0\cdot 10^{-7} \Delta
\left(\frac{E_0}{10^{51}erg}\right) \left(\frac{n}{1cm^{-3}}\right)
\left(\frac{d}{1kpc}\right)^{-2} {\rm cm}^{-2}{\rm s}^{-1}
\label{eq:fmax01}
\ee
\be
F_\gamma^{max}(>1 TeV) = 3.5\cdot 10^{-11} \Delta
\left(\frac{E_0}{10^{51}erg}\right) \left(\frac{n}{1cm^{-3}}\right)
\left(\frac{d}{1kpc}\right)^{-2} {\rm cm}^{-2}{\rm s}^{-1}
\label{eq:fmax1000}
\ee
where $\Delta$ is the fraction of the SNR kinetic energy transferred
to cosmic rays. The fluxes are a factor of 1.5 for $E_\gamma > 0.1
GeV$ and a factor of 2.6
 for $E_\gamma > 1 TeV$ lower than those of Drury et al.,(1994) the
difference being partly due to the $\pi^\circ$-production model but
mainly to the adopted form for the
energy spectrum of produced cosmic rays. It is worth remarking that
our scenario, originating in the Axford model, has no intrinsic non-linear
effects and is based on the energetics of the acceleration process.
\begin{figure}[hptb]
\begin{center}
\includegraphics[height=15cm,width=15cm]{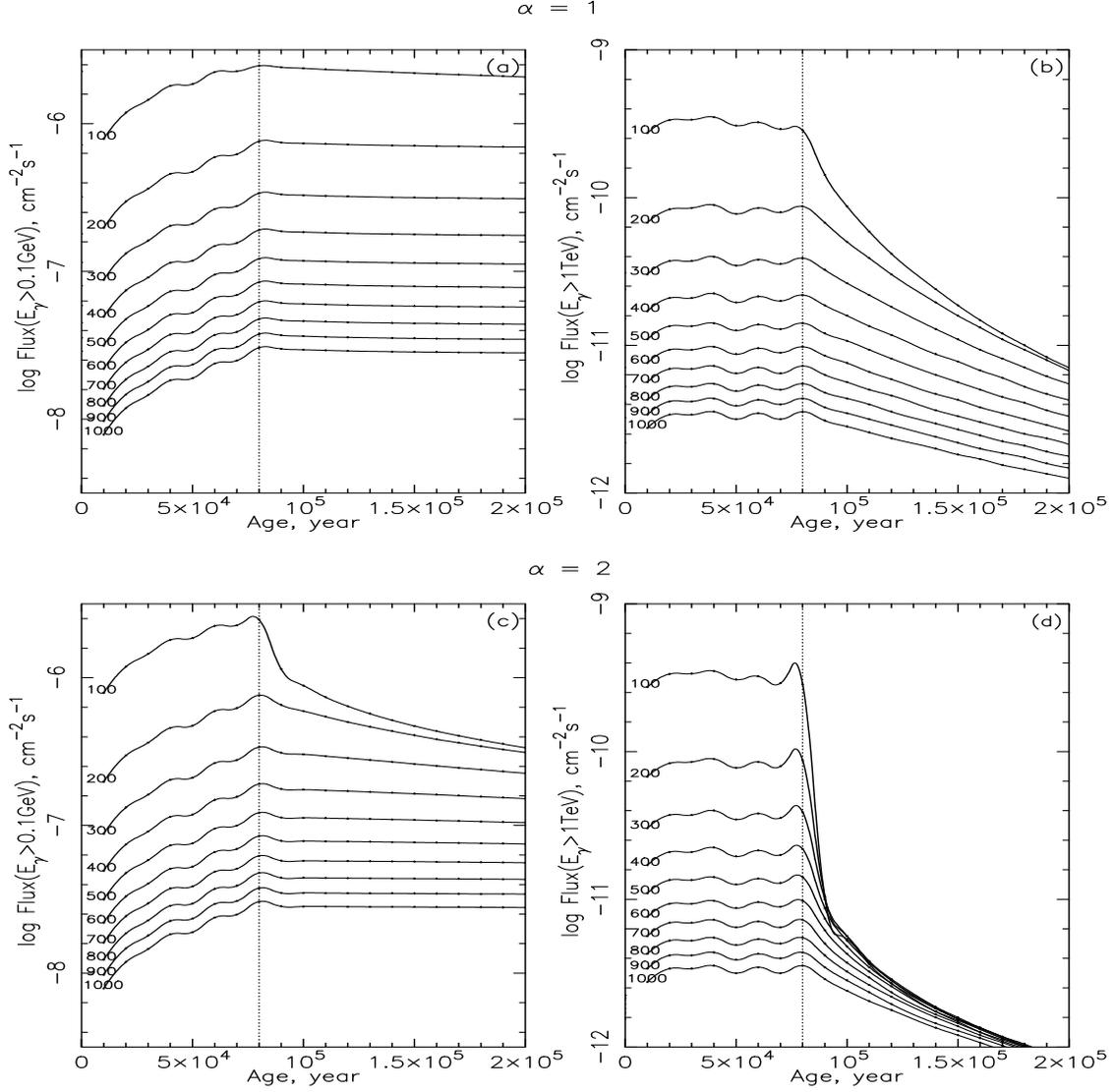}
\vspace{1cm} 
\caption{\footnotesize  Fluxes of gamma rays expected at different
distances from a SNR as a function of its age. Numbers on the curves
indicate the distance from the SNR in $pc$. (a) and (b) - for
'anomalous diffusion', (c) and (d) - for 'normal diffusion'. (a)
and (c) - for $E_\gamma > 0.1 GeV$, (b) and (d) - for $E_\gamma > 1
TeV$. Wobbles of the curves are due to the limited accuracy of the integration
 and the spline fitting of the discrete points. The different time
dependence of the flux within the remnant between the two energies
comes from eqyation (4).} 
\label{fig:vis4}
\end{center}
\end{figure}

After the expansion phase the intensity of cosmic rays decreases in a
different way depending on the gamma ray energy and the diffusion
mode. The higher energy particles diffuse and escape faster than the 
low energy ones. For the nearby SNR after a rather short time they
reach the Earth and pass it, so that a gamma ray telescope on the
Earth cannot see the gamma rays produced by those cosmic rays behind 
the Earth.

Because normal diffusion is faster at 
early times than the anomalous one the decrease of the intensity after 
the end of the expansion at $8 \cdot 10^4$ years is much faster for
$\alpha = 2$ than for $\alpha = 1$.

Nearby SNR are expected to give an extended excess of the gamma ray
intensity, i.e. extended in angular scale. If the detector has sufficient angular resolution, it can
observe the profile of the extended source. In Figures 5 and 6 we give
examples of the expected angular profiles for sources of 
different age developing by anomalous and normal diffusion. It is
seen that the profiles are even more sensitive to the energy of the gamma
quanta and the diffusion mode than are the total fluxes (~integrated
over angle~). The angular sizes of
the remnants reach tens of degrees and the usual search technique,
based on the comparison of `ON' and `OFF' runs, where both are taken
from nearby parts of the sky, is very inefficient in this case. It is 
interesting to note that, eventually, it should be possible to
determine the diffusion `law' by measuring the intensity profile for
old SNR.

A point of relevance concerns the premature escape of high energy
particles before $8\cdot 10^4$ years is reached. The frequently
fragmented nature of SNR makes this a distinct possibility and in
consequence the estimated TeV fluxes may be upper limits. However, if,
indeed, anomalous diffusion is true the core and halo will be populated by
these particles and, for distant SNR at least, the total gamma ray
flux will barely change. 
\begin{figure}[hptb]
\begin{center}
\includegraphics[height=15cm,width=15cm]{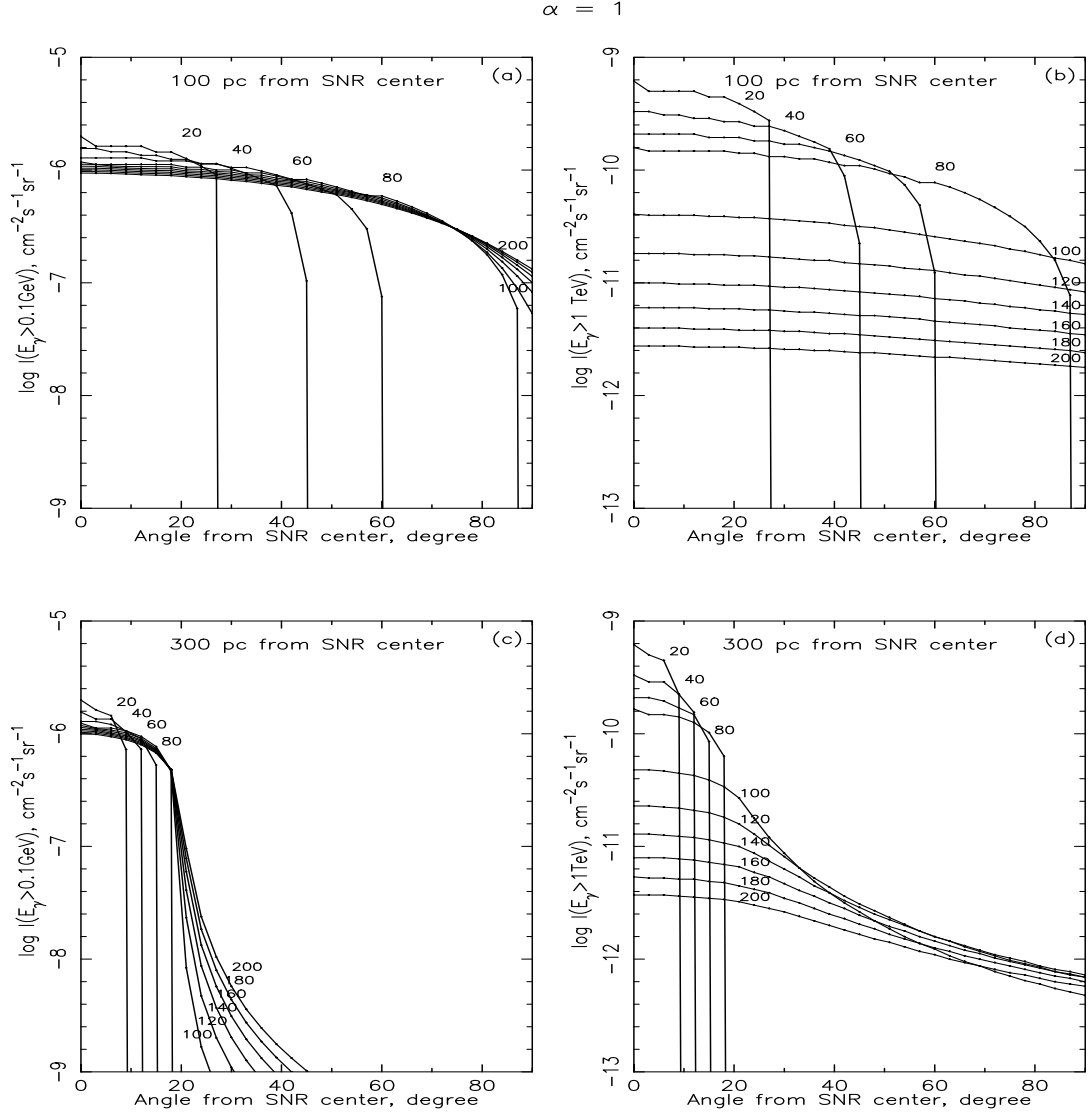}
\vspace{1cm} 
\caption{\footnotesize  Angular profile of SNR in gamma rays observed at 
distances of 100pc and 300pc from a SNR center for `anomalous'
diffusion with $\alpha$ = 1. Numbers on the curves
indicate the age of SNR in kyears: (a) and (c) - for $E_\gamma > 0.1
GeV$, (b) and (d) - for $E_\gamma > 1 TeV$. It is evident that the
core of the source will remain visible for far longer for the case of
anomalous diffusion than for normal diffusion. The number of sources
visible above a limiting intensity will therefore be greater in the
anomalous diffusion case.} 
\label{fig:vis5}
\end{center}
\end{figure}

\begin{figure}[hptb]
\begin{center}
\includegraphics[height=15cm,width=15cm]{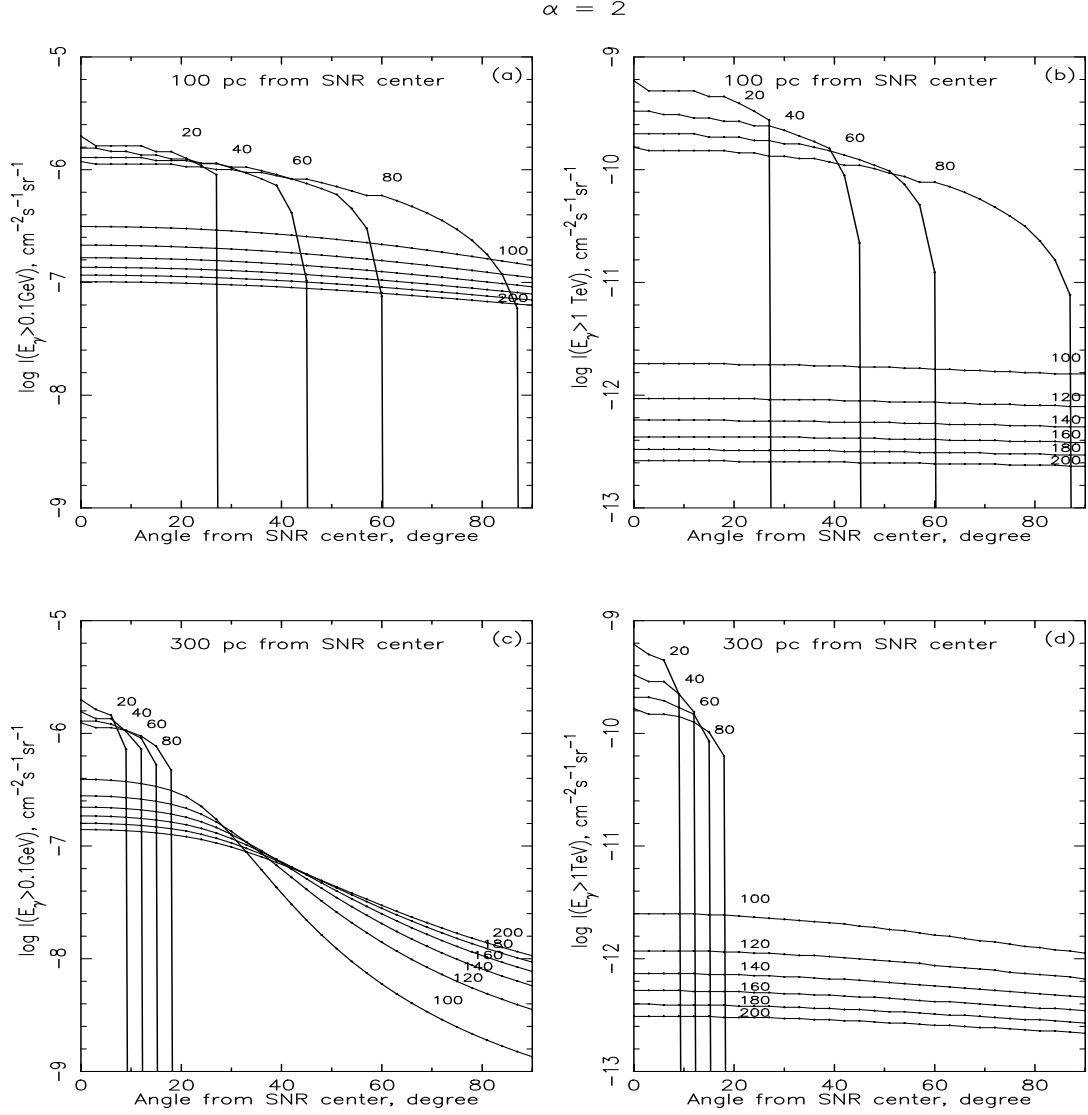}
\vspace{1cm} 
\caption{\footnotesize  Angular profile of SNR in gamma rays observed at 
distances of 100pc and 300pc from a SNR center for `normal'
diffusion with $\alpha$ = 2. Numbers on the curves
indicate the age of SNR in kyears: (a) and (c) - for $E_\gamma > 0.1
GeV$, (b) and (d) - for $E_\gamma > 1 TeV$. In this and all relevant
previous Figures the intensities relate to $p$(CR) - $H$(ISM)
collisions only, i.e. the factors referred to in \S2.3.3 are not included.} 
\label{fig:vis6}
\end{center}
\end{figure}
 
\subsection{Measured gamma ray fluxes and upper limits}
\label{sec:measured}

\subsubsection{Search for gamma rays with energies above 0.1 GeV}
{\label{sec:above100}

The third EGRET gamma ray source catalogue of Hartman et~al.
(1999), which relates to `sources' above 0.1~GeV, lists twelve
sources which had been identified by Sturner and Dermer (1995) as
being possibly associated with SNR. These sources have median
fluxes (~the measured flux varied somewhat from one viewing period
to another for the same source~) in the range
(3-12)$\times10^{-7}\ {\rm cm}^{-2}{\rm s}^{-1}$, with an
overall mean flux of $6.4\times10^{-7}\ {\rm cm}^{-2}{\rm s}^{-1}$.

A later paper by Sturner et~al. (1996) has reduced the number to
five high confidence coincidences and two marginal
coincidences. These associations are listed in Table~1. Three of
those in the new list were in the old list (plus the two marginal
cases). There are two new ones and six of the old ones have
disappeared. The three common ones have a mean flux of
$8.7\times10^{-7}\ {\rm cm}^{-2}{\rm s}^{-1}$, compared with
$6.4\times10^{-7}\ {\rm cm}^{-2}{\rm s}^{-1}$ for the whole earlier
set and are thus, presumably, more reliable.

The extent to which the SNR under consideration is associated with 
a nearby molecular
cloud is indicated in the Table; by `interacting' is meant such an
association.

As the authors imply, the origin of the gamma rays in terms of CR
acceleration by the SNR followed by interaction in the molecular
cloud is only one possibility. Pulsars, or just the gas
enhancements coupled with the ambient CR intensity, are also
possibilities.

The uncertainty is also referred to by Torres et~al. (2002) who
include, in their analysis, new CO-observations (and implied
molecular hydrogen masses). The importance of molecular clouds as
targets has also been well brought out in a very recent paper by
Torres et al. (~2003~). These authors point out that even distant SNR
(~further than 5 kpc~) may be detectable as gamma ray sources if
nearby clouds are sufficiently massive. They present evidence for 5
`coinciding pairs of 3EG sources and SNR', for which there is evidence
of nearby large molecular clouds.

The conclusion to be drawn at this stage is that there are
probably coincidences of young (~less than about $10^4$~y~) SNR with
gamma ray sources, but there is no certainty, yet. If the
coincidences {\em are} genuine then the fluxes (above 0.1~GeV)
are of order $(6-9)\times10^{-7}\ {\rm cm}^{-2}{\rm s}^{-1}$. Perhaps
the best example is $\gamma$ Cygni which, with a median flux of
$1.2\times10^{-6}\ {\rm cm}^{-2}{\rm s}^{-1}$, is the nearest to us, at a
distance of $\sim1.5$~kpc.

\subsubsection{The results for the nearby `Loops'}
\label{sec:nearbyloops}

There were early claims that `Loop I' had been detected in this
energy band. Thus, Lebrun and Paul (1985) and Bhat et~al. (1985)
claimed excess fluxes from the direction of this feature. The
latter analysis related to results from both the SAS$\,$II and
COSB satellites. Rogers and Wolfendale (1987) went further and
claimed excesses for Loop~III and the Vela region. The work by
Wolfendale and Zhang (1994) and Osborne et~al. (1995) should also
be mentioned. These authors examined data from the Compton Gamma
Ray Observatory, specifically from the EGRET instrument
and found gamma ray excesses from the `ridges' of a number of SNR.
Loop~I was examined in detail by Osborne et~al. and a strong case
made for the presence of CR acceleration.

The best estimate of the excess gamma ray flux from Loop~I is
$F_\gamma(>0.1\ {\rm GeV}) = (1.2\pm0.3)\times10^{-5}\ {\rm
cm}^{-2}{\rm s}^{-1}$. We have reexamined the results from all the
satellites and endeavoured to find the mean intensity through the
centre of the remnant, as distinct from the overall flux, for reasons
that will become clear later. The value is $(5 \pm 2) \cdot 10^{-6}
{\rm cm}^{-2}{\rm s}^{-1}{\rm sr}^{-1}$.

Information on the relative contribution of electrons and protons 
(~and heavier nuclei~) was given by Osborne et al.(1995). These workers made
an analysis of the spectral shape far from the ridge, and nearby,
and found similar shapes. Now Strong and Mattox (1996) showed the
presence of the `pion-peak' in the general diffuse flux, i.e. a
majority of protons, so that the Loop I excess can also be explained by
protons. 

A consequence of the SN explosions which created our Local
Superbubble is an environment enriched by nuclei. It is known that 
their abundance in cosmic rays grows with the energy. According to
our 'Single Source Model' these nuclei are responsible for the fine
structure of the energy spectrum at the highest PeV energies with oxygen as the
dominant element at the knee. The higher gamma ray emissivity of heavy nuclei
compared with the protons can, therefore, in principle contribute to
the observed flux of gamma rays from Loop I at higher energies.

Loop~I is, in our view, a very likely source of
excess gamma rays. A comparison with our expectation will be given later.

\subsection{Search for gamma ray-SNR associations in the TeV energy region}
\label{sec:searches}

\subsubsection{General comments}

The advent of Cherenkov radiation telescopes, which detect photons
from secondary electrons in the upper levels of the atmosphere, has led to
advances at TeV gamma ray energies. In terms of the physics to be
studied, there is the usual advantage of going to higher
energies–--with the possibility of surprises. There is also the
(hoped for) likelihood of the initiating particles being protons
(and heavier nuclei) rather than electrons.

In terms of detectability, the relative background due to cosmic
rays interacting with the Galactic ISM is reduced. This is because
SNR have CR spectra on emergence of the form
$E^{-2.15}$, whereas the ambient CR have an exponent of about 2.7;
the difference arises because of the energy-dependent Galactic
escape.

\subsubsection{SNR beyond 1 kpc}
\label{sec:TeV-SNR}

An important search has been made by the HEGRA-IACT group
(Lampeitl et~al., 2001, Aharonian et al., 2002~), relating to signals
from 63 SNR on a quarter of the Galactic Plane ($-2^\circ < l < 85^\circ,
-1.7^\circ < b < 1.7^\circ$); none was detected. It is important to
appreciate, however, that only 19 are situated at known distances
(Green, 2000). Of the 19, the nearest is at 2.7~kpc
(G023.3$-00.3$), and the mean distance is 5.2~kpc. The remaining
44 SNR are probably beyond 6~kpc, in view of their low
radio fluxes. The upper limit for the SNR population is at the level
of $1.2\cdot 10^{-12} {\rm cm}^{-2}{\rm s}^{-1}$ for $E_\gamma > 1 TeV$. 
   In total, just 5 distant SNR have been detected in the TeV energy
region, 2 of them are plerions (~Crab and PSR 1706-44~), which we
shall not analyse here. The other 3 detected are of the shell type as follows: 
\begin{itemize}
\item[(i)] Cas A (3C461). This is the remnant from the SN which
exploded in 1680; it is 2.8~kpc away, has a radius of 2.5~arc$\,$min
and a radio flux of 2700~Jy.
The observed flux above 1 TeV (HEGRA: Aharonian et al., 2001) is
$6\times10^{-13}\ {\rm cm}^{-2}{\rm s}^{-1}$.
\item[(ii)] SN$\,$1006 (G 327.6+14.6). This is the remnant of the
SN which exploded in 1006. It is $\sim1.8$~kpc away and has a
radius of $\sim30$~arc$\,$min. The radio flux at 1~GHz is 19~Jy.
The flux observed above 1 TeV (from part of the shell: CANGAROO: Hara et~al., 
2001) is $6\times10^{-12}\ {\rm cm}^{-2}{\rm s}^{-1}$.
\item[(iii)] SNR G348.5+0.0/348.5+0.1/347.3-0.5 associated with
RX J1713.7-3946. This is an example of an extended source, which
emits gamma quanta thought to be produced in the interactions of cosmic rays
accelerated by the SNR shell with the nearby dense molecular clouds.
The distance and the age of the source are under discussion and vary
from 1 kpc with 2 kyears to 6 kpc with more than 10 kyears. The observed 
flux above 1 TeV is $\sim 1\times10^{-11}\ {\rm cm}^{-2}{\rm s}^{-1}$ 
(CANGAROO: ~Enomoto et al., 2002~). 
\end{itemize}

\subsubsection{SNR within 1 kpc}

Green describes 5 SNR in this category, including the Cygnus
Loop, Monoceros and Vela. Details are given in Table~2. Only
Vela, which is a plerion type SNR, has been detected. Among nearby SNR
found in X-rays the diffuse Monogem Ring should be mentioned
(~Plucinsky et al., 1996~), this object is located near Monoceros
Nebula. In spite of the fact that it has not been detected in gamma
rays we mention
it as one of the possible candidates for our nearby (~$\sim$ 300 pc ) and
recent ( $\sim 9\cdot10^4$ years~) `Single Source'. The expectation 
values will be considered later.

\subsection{The sensitivity problem and the minimum detectable flux}
 \label{sec:sensitivity}

It is necessary to examine the general question of the sensitivity
of contemporary detectors to searches for gamma rays
(~$E_\gamma>0.1$ GeV and $>1$ TeV~) from SNR. Clearly, the
sensitivity will be
a function of the search method and of many parameters, most notably 
the background (~which
will be a maximum in the Galactic Plane and towards the Inner
Galaxy~) and the angular size of the source. Insofar as the majority of
SNR are in the Galactic Plane the former will be more important, but in
the case of Loop~I, and our `Single Source', the latter will also
be of significance. The bigger the source the more accurately does the
background need to be known because an $l,b$-dependent background
makes the technique of `ON source - OFF source' much more difficult. 
The characteristics of the gamma ray telescope: area and angular 
resolution are also of course of prime importance.

In view of what has been said already, we consider that Figure \ref{fig:vis7}
gives a good representation of the current limiting fluxes. As
given, the values are averages (over the Galactic Plane); there is
an uncertainty of a factor of about 2 to allow for the variable
background. This, then, represents the contemporary limit to the identification
of there being a significant gamma ray excess in the direction of
a known SNR in, or near, the Galactic Plane. 

Turning to the interpretation of observed excesses, there are further
uncertainties. The excess {\em could} be due to
an underestimate in the amount of gas in the vicinity. Li and
Wolfendale (1981) examined this aspect for the 2CG catalogue of
COS$\,$B sources and in that case showed that only half the
sources were `genuine' and not just due to the ambient,
near-constant CR flux irradiating molecular clouds. Houston and
Wolfendale (1984) estimated that a `genuine' source of flux above
0.1~GeV of $10^{-6}\ {\rm cm}^{-2}{\rm s}^{-1}$ has a detection
probability of less than 10\% at $l=30^\circ$ and still only
$\simeq50\%$ at $l=120^\circ$. Although the subject has moved on
in the intervening period it is difficult to imagine that the
average fluxes given in Figure \ref{fig:vis7} are too high.

As with the situation for low energy gamma rays, it is important
to consider the level at which the TeV gamma ray sources might be
visible. Although this level is dependent on the characteristics
of individual detection arrays (~e.g. on the size of the mirror and the
angular bin size allowable~) there will be the same general feature as
at low energies: the threshold flux will rise
with increasing size of the object, viz. weak extended objects will be 
more difficult to see than sharp localised ones. Inspection of a
collection of arrays gives the line shown in Figure \ref{fig:vis7}.
\begin{figure}[hpt]
\begin{center}
\includegraphics[height=9.5cm,width=15cm]{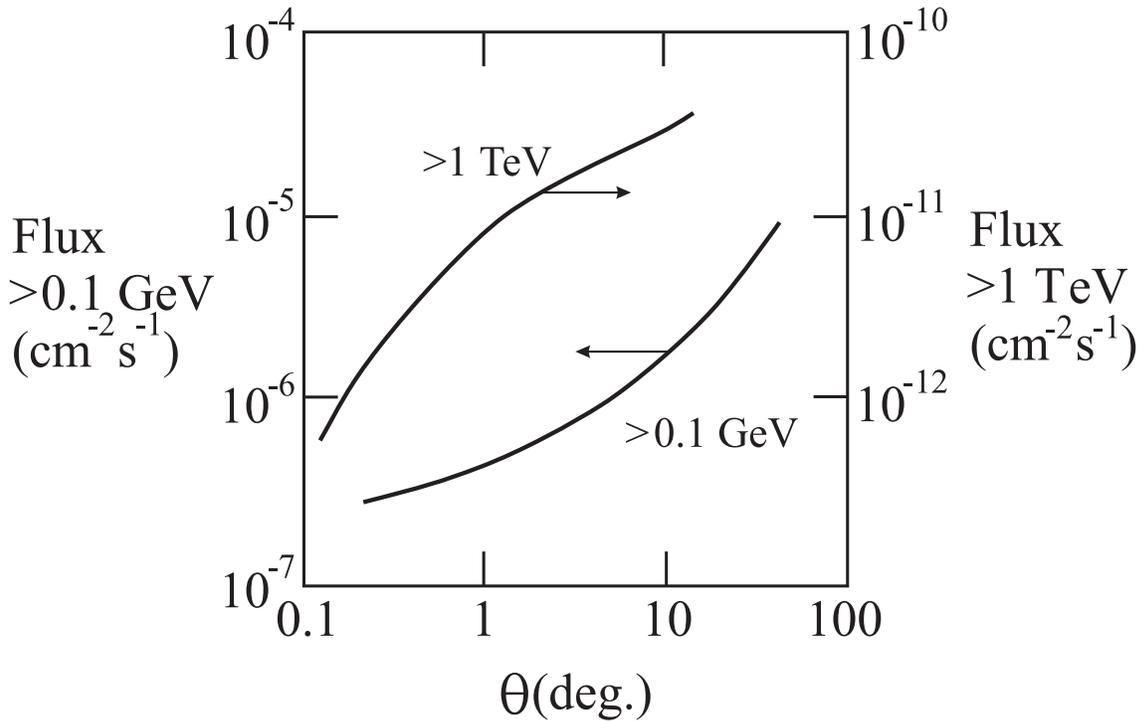}
\vspace{1cm} 
\caption{\footnotesize Minimum detectable fluxes of gamma rays with
energy above 0.1 GeV (~lower curve~) and above 1 TeV (~upper
curve~) as a function of the angular size of the source.} 
\label{fig:vis7}
\end{center}
\end{figure}

\section{The density of the ISM in the vicinity of SNR and the CR
energy content}

\subsection{General remarks}

The density of the ISM is a key parameter in our calculations
insofar as the ISM provides the target material for gamma ray
production. Although the canonical mean density of the ISM is
1~H-atom$\cdot$cm$^{-3}$, and this has been used as our datum in the
calculations, it is virtually certain that (~in the absence of
molecular gas~) the relevant value will
almost always be lower here. The basic reasons are two-fold:
\begin{itemize}
\item[(i)] SN often occur in groups so that there is a good chance
of the SN under consideration having exploded in a `bubble' caused
by one or more previous SN.
\item[(ii)] Type~II SN, which are thought to be responsible for CR
acceleration, are caused by very massive stars (often O-stars and
Wolf-Rayet stars) and these stars give rise to strong winds before
the SN itself. The winds displace much of the gas surrounding the
star so that the remnant expands into its parent star's evacuated
surroundings. For very young SNR, however, (~eg. Berezhko et al.,
2003~) the previous wind may cause an {\em increase} in gas density. 
Details are given in what follows.
\end{itemize}
It will be appreciated that insofar as matter is
conserved, there will be a pile-up of IS gas and the
particle-irradiation of this gas may need to be
considered. Furthermore, there is the ever-present possibility of the
presence of molecular gas.

\subsection{SN associations} \label{sec:associations}

Many of the massive stars responsible for SN are found in
associations. Parizot et al.,(2001) quote 90\% for this fraction,
whereas Wallace et al.,(1994) and Ferri\`{e}re, (2001) quote `about
half' for the catalogued O-stars and their association with
clusters. These massive objects have energetic stellar winds
which inject as much as $10^{51}$~ergs into the ISM over the
stellar lifetime, i.e., as much  as injected by the eventual SN
itself. Arnal and Mirabel (1991), Lozinskaya (1991) and others have
reported bubbles of up to 50--100~pc in
diameter. Heiles (1987, 1990) has made the point that since many
of the clustered O- and B-stars become SN `together', within a short
space of time (they are of almost the same age), they will form
`superbubbles' several hundred pc in diameter.

Loop~I is almost certainly an example of a close succession of SN in a
confined space and SNR G78.2+2.1, discussed later in connection with
Gamma Cygni source, is another example of an SNR which is a member of an 
association.

\subsection{Excavation of the ISM by an individual SN}

For a single, isolated, SN the precursor star will reduce the
local ISM density if, as is common (see
Section~\ref{sec:associations}), the stellar wind had a high energy
content. Thus, many Type~II SNR will expand into `low
density' regions of the ISM. The effect of this expulsion of ISM
gas from the region is presumably not serious for the distant sources
from the standpoint of overall, eventual, gamma ray production in the
sense that because the mass of gas is
conserved the gamma ray emission is simply transferred from
within a near-uniform sphere to a spherical shell which for the
distant sources is still within the field of view of the
detector. However, for young SN, the measured gamma ray flux can be low
even for this case if, as is likely, the SNR shock does not reach the
compressed-gas region. For the nearby sources the effect of the ISM
excavation should be substantial, because a large portion of the
target mass will be outside of the fiducial volume of the detector. SNR Cas
A, discussed in Section \ref{sec:TeV-SNR} and later in Section
\ref{sec:detTeV}, is an example of such an SN which exploded into a wind
bubble created by the progenitor star although Berezhko et
al. (~2003~) argue that the gas density may be high because `part of
the slow red supergiant wind of the SN progenitor has been swept up
into a dense shell'. SN 1006 and RX J1713.7-3946
also reside in a low density environment.

\subsection{The presence of molecular clouds}
\label{sec:molclouds}

A complication is the presence of molecular gas which will be target
material for the primary particles accelerated within the remnant
(~and, of course, will be a target for the ambient CR in the ISM in
general, eg. Houston and Wolfendale, 1981~). The situation is not
straightforward for a number of reasons, as follows: 
\begin{itemize}
\item[(i)] The presence of molecular gas, mainly hydrogen, is
manifested by the observation of the CO-line and there are many maps
of CO-emission. The overlap of such contours with an optically
detected SNR does not guarantee the presence of H$_2$ {\em within} the
remnant, because of line of sight effects, however. Indeed, only IC443
seems to be a clear example of a significant interaction between SNR
and a molecular cloud (~MC~).
\item[(ii)] Strong shocks dissociate CO and thus may `remove the
evidence'. The dissociated H$_2$ may or may not be seen against the
foreground and background material.
\item[(iii)] The interaction of an SNR shock with a MC is a matter of
great complexity, not least because a MC itself is a complex system,
composing as it does a mixture of dense clumps and less-dense
inter-clump material.

Concerning the interaction of a shock with a clump (~density several
hundred atoms cm$^{-3}$~), the shock speed is reduced considerably
and particles may not be allowed into much of the cloud by the time of
observation even though the rest of the shock has moved past the
cloud. This factor might be the limiting one for `young' SNR (~age
$\lesim 10^4$ years~). For example, in a cloud of mean density $3\cdot
10^2 {\rm cm}^{-3}$ compared with the standard  $3\cdot 10^{-3} {\rm
cm}^{-3}$, the density ratio is $10^5$ and, since the shock velocity
is proportional to $n^{-0.2}$ (~Dyson and Williams, 1980~), the speed
will be reduced by 10. For a remnant of age $1.35\cdot 10^3$ years
(~$\gamma$-Cygni - see later~), the shock velocity will be reduced 
to $\sim 500 {\rm km} {\rm s}^{-1}$, i.e. 0.5 pc per 10$^3$ years.

A further point of importance is that energy is taken from the shock
and given to the molecular material, thereby reducing the amount
available for CR acceleration.

\item[(iv)] Diffusion of particles into the cloud from the surrounding
previously shocked medium may be slow. For example, for 2 GeV
particles (~the precursors of gamma rays above 0.1 GeV~) even with the
`normal' diffusion coefficient used by us in the ISM in general
(~$D\sim 10^{28} {\rm cm}^2{\rm s}^{-1}$~) the {\em rms} penetration
will only be ~$\sim 6$ pc in 10$^3$ years. If the diffusion
coefficient is much smaller in regions of high density - as would be
expected for reasons of magnetic field compression - then the
penetration will be even slower. In this connection it can be remarked
that in their SNR shock - CR acceleration paper, Berezhko et al., (1996)
assumed that the diffusion coefficient is inversely proportional to
density. For a cloud with mean density $\sim 100 {\rm cm}^{-3}$, the
{\em rms} penetration distance will therefore be reduced by an order of
magnitude. 

A relevant point concerns the ionization density in dense MC
(~Chevalier, 1999~), however. This author argues that CR may, in fact,
propagate rapidly if the Alfv\'{e}n waves are heavily
damped. Nevertheless, the energy transfer problem, i.e. reduced
fraction available for CR, cannot be circumvented.
\item[(v)] The role of the inter-clump material appears
crucial. Following Blitz, (1980) we note that although the fraction of
the mass carried by this `low' density material (~$\langle n \rangle
\sim 11 {\rm cm}^{-3}$~) is low (~$\sim$ 25\%~) the fractional volume
is high - $\sim$ 92\%. Chevalier, (1999) and Bykov et al., (2000) have
argued for the generation of non-thermal radiation by magnetic field
compression in this material. Electron-bremsstrahlung can be
important, here (~see later~).
\end{itemize}
A further complication is that the shock may break up the cloud
somewhat and thereby increase the mean rate of
penetration. Nevertheless, the effective mass of a MC
within, or adjacent to, a remnant will usually be much less than its
inferred mass, and for $\pi^\circ$-production, at least, the yield will be
reduced.

The effect of the above will be particularly important for
searches for gamma ray sources associated with known SNR-MC
associations and particularly for those SNR which are young.

\subsection{The mean gas density to adopt}

The crucial question is the nature of the SNR. If it has not been
recorded as such by optical means, then, if the `object' is large, the
effective density may be quite high, because the piled-up gas may well
be included. On the other hand, searches for
gamma ray emission from known sources - the usual situation - in
general deals with young SNR where the shock has not reached the piled-up gas
and the mean density of gas is low, unless there is nearby molecular
gas, or unless the precursor has provided considerable local gas.

Taking the factors listed above into account, it appears
reasonable to use a mean value equal to about twice that in the Local
Bubble, this latter being $~3-4\times10^{-3}\ {\rm cm}^{-3}$
(~see \S2.1~) for optically detected SNR for which there is no
known associated molecular material. Thus, we adopt 
 $\langle n \rangle = 10^{-2} \ {\rm cm}^{-3}$, but bear in mind that
there will be big
fluctuations in this value, not least if the SNR abuts or contains a molecular
cloud, or has provided considerable gas from the precursor star. In
fact, in the absence of molecular gas the distribution in
densities will be nearer
bimodal, with $\sim50$\% having $n\sim10^{-3}\ {\rm cm}^{-3}$ (~young SN
in a cluster~) and 50\% with $n\sim10^{-1}\ {\rm cm}^{-3}$ (~a
single isolated comparatively old SN~).

\subsection{Energy taken by cosmic rays} \label{sec:energytaken}

Although we adopt $10^{50}$~ergs as the energy going into cosmic
rays, it can be different. Examining the work of Berezhko et~al.
(1996) we note that this quantity can in fact be as high as 80\% of the
total, viz. $8\times10^{50}$~erg. These authors give 80\% and
$\sim25\%$ for the CR energy for an initial Mach number of 33 and injection rates
$10^{-2}$ and $10^{-4}$ respectively.

\section{Comparison of observations with our estimates}

\subsection{Gamma ray energies above 0.1~GeV in general}
\label{sec:gen01}

For SNR beyond 1 kpc, viz. less than about $1^\circ$ in radius,
Figure~\ref{fig:vis7} shows that the present limiting observational
flux is about
$5\times10^{-7}\ {\rm cm}^{-2}{\rm s}^{-1}$. Turning now to
Figure~\ref{fig:vis4}a, which is for our preferred propagation model
($\alpha=1$), the predicted flux is 
$\sim 3\times10^{-8}\ {\rm cm}^{-2}{\rm s}^{-1}$ for $pp$ interactions
and $n = 1 \ {\rm cm}^{-3}$.
For all CR-ISM interactions the value rises to 
$7\times10^{-8}\ {\rm cm}^{-2}{\rm s}^{-1}$. This is already below
the observed limit of $5\times10^{-7}\ {\rm cm}^{-2}{\rm s}^{-1}$.
For the 50\% of sources with $n\sim10^{-1}{\rm cm}^{-3}$ the predicted
flux is $7\times10^{-9}\ {\rm cm}^{-2}{\rm s}^{-1}$, i.e. only
$\sim$1.4\% of this limit. Even with an 8-fold increase of the total CR
energy content (see Section~\ref{sec:energytaken}) the predictions are
still an order of magnitude below the observed limit. It is,
therefore, not surprising that many of these SNR are not seen.

For sources within 1 kpc, although the predicted fluxes are
bigger, because of their increased angular diameters, the threshold
fluxes for the sources will also be  bigger. As an example, an SNR at
300 pc has a maximum predicted flux of $\sim 3\times10^{-7}\ 
{\rm cm}^{-2}{\rm s}^{-1}$ for $n = 1 {\rm cm}^{-3}$ (~Figure
\ref{fig:vis4}a~), for $pp$, i.e. 
$\sim 7\cdot10^{-7}{\rm cm}^{-2}{\rm s}^{-1}$ for CR-ISM
interactions. With  $n \sim 10^{-1} {\rm cm}^{-3}$ - the highest
probable density in the absence of molecular clouds within the
remnant - the predicted flux is $\sim 7\cdot 10^{-8}{\rm cm}^{-2}{\rm
s}^{-1}$. At 300 pc, the angular radius of SNR at the end of the
expansion phase is about 20$^\circ$ and Figure \ref{fig:vis7}
indicates an observed limit of 
$\sim 3\cdot 10^{-6}{\rm cm}^{-2}{\rm s}^{-1}$. Again, even with an
eightfold increase in energy content, the SNR will still be
non-detectable in low energy gamma rays.

\subsection{The detected sub-GeV gamma ray sources}

\subsubsection{The likely SNR---gamma ray source coincidences}
\label{sec:likelySNR}

Table 1 gives a list of these coincidences ( see also Section 
\ref{sec:above100}). At the indicated distances and ages, the expected
flux will be
$\lesim 3\cdot10^{-9} {\rm cm}^{-2}{\rm s}^{-1}$ for $n = 1 {\rm
cm}^{-3}$, the low flux being due to both distance and youth (~Figure
\ref{fig:vis4}a~). For  $n = 10^{-1} {\rm cm}^{-3}$ the result is 
$\lesim 3 \cdot 10^{-10} {\rm cm}^{-2}{\rm s}^{-1}$, several orders
less than observation (~$\approx 10^{-6} {\rm cm}^{-2}{\rm s}^{-1}$~).
Presumably, the answer lies in some way with the accompanying `molecular cloud'
although the increased effective density required ($\simeq 300 {\rm
cm}^{-3}$~) is very high, implying that most of the generated particles
traverse the cloud if CR - gas nucleus collisions are in fact
responsible, or more likely, that another process is more important.
 A `case history' will examine this aspect. 

Now some more details about the most important of the detected
sources.
 
\subsubsection{Loop I}

The flux above 0.1~GeV is estimated by us as $\sim10^{-5}\ {\rm
cm}^{-2}{\rm s}^{-1}$ (~see Section \ref{sec:nearbyloops}~). 
The estimates of the distance to the Loop I
center and of its age are quite uncertain: $130 \pm 75$pc and 
of order $10^5$ years, respectively. However, even with these uncertainties
it is possible to say that for a {\em single} source the expected flux 
from our Figures 4a and 4c
is between $0.6 \cdot 10^{-6}$ and $2.4 \cdot 10^{-6} 
{\rm cm}^{-2}{\rm s}^{-1}$,
i.e. is lower by the factor of 4 even for the standard matter density
of $n \approx 1 {\rm cm}^{-3}$. Because Loop I is thought to be
responsible for the formation of our Local Superbubble, the actual
matter density in it is reported as $3\times10^{-3}\ {\rm cm}^{-3}$ and the
expected flux should be reduced to $\sim 10^{-8} {\rm cm}^{-2}{\rm
s}^{-1}$ for a single source.  Even allowing for the fact that Loop~I 
almost certainly results
from many ($\lesim 10$) SN over the past 1~My and allowing for an
increase in CR energy, as well as with the possibility of enriching the
mass, the discrepancy is still large. 

A possible way out is the following. Although Loop I is in
the HISM the measurements referred to in 
Section~\ref{sec:nearbyloops} show that the ridges stand out, i.e. the
shock has reached the `piled-up' gas. Furthermore, Frisch, (1997) has
pointed out that there appear to be two molecular clouds in the
remnant. These clouds are about 20 pc in radius and since the age of
the remnant is large (~$\sim 10^5$ years~) there has been time for the
clouds to have been penetrated by the CR. The
corresponding mean gas density is $10^{-1} {\rm cm}^{-3}$ (~our higher
component value, by chance~), and the predicted intensity in the
centre $10^{-6} {\rm cm}^{-2}{\rm s}^{-1}{\rm sr}^{-1}$, i.e. a factor
(~5$\pm$2~) short of the measurement. It is possible to
achieve  agreement by increasing the CR energy yield (~the individual
CR particle energies are
too low for it to be allowable to increase the mean mass~),
furthermore, several SN over the last Megayear were probably 
responsible for the remnant, a claim in accordance with that of
others. Agreement follows.

\subsubsection{Gamma-Gygni}

This source is important because it is often used as an example of a
strong 'GeV source' which on the other hand is a very weak 'TeV
source', which means that it is claimed that it cannot accelerate 
protons and nuclei up
to TeV energies (~Prosch et al., 1996; Plaga, 2001~).
The situation here is the following. The observed
flux of gamma rays above 0.1~GeV is 
$1.2\times 10^{-6}{\rm cm}^{-2}{\rm s}^{-1}$ for the source. 
There is an associated SNR, G78.2+2.1, which is a member of the Cyg OB9 association and
if for this source alone $n \sim 10^{-3}{\rm cm}^{-3}$
the expected flux of gamma rays might be as low as 
$10^{-11}{\rm cm}^{-2}{\rm s}^{-1}$. However, if the nearby molecular
cloud Cong 8 is indeed physically coupled with the radio bright region
DR 4, which is the evidence for the interaction of the accelerated
electrons with the magnetic field of the cloud, then taking into account
the parameters of the cloud (~$n \sim 2700{\rm cm}^{-3}$ and 0.05 for
the fraction of the volume which it occupies in the SNR~), Prosch et
al. were able to explain the observed gamma ray flux on the basis
of its hadronic origin.  

Some further comments are needed, however. It is, indeed, true, that
there appears to be evidence for the impact of the SNR on Cong 8 by
way of distortion of the radio contours but the fraction of the mass
within the remnant appears to be less than 5\%. Furthermore, Prosch et
al. assume a mass for Cong 8 of $\sim 7\cdot 10^4 M_\odot$ whereas the
truth seems to be much less (~$\sim 6\cdot 10^3 M_\odot$, following
Pollock, 1985~). Indeed Cong 8 is a small part of the Cyg OB9 complex
which, itself, only has a mass of $\sim 10^5 M_\odot$. There is also a
problem of the age of the remnant. Various values have been quoted, as
high as $2\cdot 10^4$ years. Adopting the angular size of the
optically `seen' SNR (~Green, 2000~) and a distance of 1.5 kpc we find
a radius of 13 pc and thus an age of $1.35\cdot 10^3$ years only, for
an ISM density of 1 atom cm$^{-3}$. 
The age is important for two reasons: the extent of shock penetration and
the ambient CR intensity at the time of observation. As the age
increases the degree of penetration increases but the CR
intensity falls. There is thus a measure of compensation. 

Assuming, for the moment, complete penetration of the cloud, the
expected flux (~above 0.1 GeV~) is given by
\be
F_\gamma = 1.3\cdot 10^{-7} M_5 (\frac{q_\gamma}{4\pi})_{26}/d_k^2 {\rm cm}^{-2}{\rm s}^{-1}
\ee
where $M_5$ is the mass in units of $10^5 M_\odot$,
$(\frac{q_\gamma}{4\pi})_{26}$ is the emissivity in units of $10^{-26}$
atom$^{-1}{\rm s}^{-1}$ and $d_k$ is the distance in kpc.
For the situation indicated above (~$1.35\cdot 10^3$ years, etc.~)
inserting a factor $f$ for the fraction of the cloud penetrated
(~f$<$1~), 
the gamma ray flux from Cong 8 will be 
\be
F_{\gamma,C} = f\cdot 3.2\cdot 10^{-8} {\rm cm}^{-2}{\rm s}^{-1}
\ee
With an assumed density within the remnant of 1 cm$^{-3}$ (~it is in a
complex region with much material about~) the contribution from the
rest of remnant is
\be
F_{\gamma,1} = 1.1\cdot 10^{-9} {\rm cm}^{-2}{\rm s}^{-1}
\ee
The total is thus $\simeq 3.3\cdot 10^{-8}{\rm cm}^{-2}{\rm s}^{-1}$ for
the limiting case of $f = 1$ (~in fact we show later that $f \sim
0.01$~).
The result is already less than the `observed' flux of $1.2\cdot 10^{-6}{\rm
cm}^{-2}{\rm s}^{-1}$ by a significant factor.

It is true that our figure is dependent on SNR age but greater age
would give a lower predicted intensity.

To conclude about low energy gamma rays for Gamma Cygni, it seems that
the contribution from SNR - accelerated protons interacting locally is
too small to explain the observed flux even if all the CR can
penetrate the molecular cloud. It is here that the proposal of
Chevalier, (1999) and Bykov et al., (2000) involving electrons is very
attractive.

The TeV region for this source can also be considered here. At these
energies only upper limits were obtained.  
From the results of HEGRA IACT (~He$\ss$ et al., 1997~) and HEGRA AIROBICC (~Prosch
et al., 1996~) systems it can be concluded that the gamma ray flux
above 1 TeV is less than $\sim 3\cdot 10^{-12}{\rm cm}^{-2}{\rm s}^{-1}$.
The predicted flux about 1 TeV from the present work is $3.5\cdot
10^{-4}$ times the flux above 0.1 GeV, i.e. $F_\gamma = f\cdot
1.1\cdot 10^{-11}{\rm cm}^{-2}{\rm s}^{-1}$. With $f = 1$, this flux
is higher than the observed upper limit by a factor 4, but the
slowness of the shock front means that $f$ will be much less than
1. Specifically, with a cloud size of 10 pc the time taken to cross it
is $\sim 2\cdot 10^4$ years (~see Section \ref{sec:molclouds}~) and, 
insofar as only about 200 years are
available (~from inspection of the geometry of the system~) $f \approx
1$\%.

Thus, we predict $F_\gamma \sim 1.1\cdot 10^{-13} {\rm cm}^{-2}{\rm
s}^{-1}$, `comfortably' below the observed upper limit.

\subsubsection{The nearby SNR}

Turning to Table 2, only Monoceros is a feature in Sturner and Dermer's
list, and here the flux is measured to be $\sim 0.4\cdot10^{-6} {\rm
cm}^{-2}{\rm s}^{-1}$ above 0.1 GeV. At a distance of $\sim$ 1 kpc and
age $\sim 10^4$ years (~radius 30 pc~), Figure \ref{fig:vis4}a
predicts $\sim 3\cdot 10^{-9} {\rm cm}^{-2}{\rm s}^{-1}$ for $n = 1
{\rm cm}^{-3}$, viz. 3 orders short. The effective density  needed is
even higher than above (~Section \ref{sec:likelySNR}~): $\sim 1000~{\rm
cm}^{-3}$ and even with an increased CR energy output and heavy
nuclei, considerable penetration of the (~known~) molecular cloud is
needed.

\subsection{Gamma ray energies above 1~TeV in general}

Arguments similar to those presented in Section \ref{sec:gen01} pertain here.
For the cluster SNR, with $n \sim 10^{-3} {\rm cm}^{-3}$, seen as the
source of the angular size $\sim 1^\circ$, our maximum predicted
flux (Figure ~\ref{fig:vis4}b) is $\sim 3\times10^{-15}\ {\rm cm}^{-2}{\rm
s}^{-1}$ for $pp$ and $\sim 7.5\times10^{-15}\ {\rm cm}^{-2}{\rm
s}^{-1}$ for CR-ISM interactions, to be compared with a minimum
detectable flux of 
$\sim10^{-11}\ {\rm cm}^{-2}{\rm s}^{-1}$ (~Figure~\ref{fig:vis7}~). 
For isolated sources of a similar size with
$n \sim 10^{-1}{\rm cm}^{-3}$ there is still a discrepancy by a factor
of 13. The larger angular size of nearby SNR requires
a smaller distance to be seen, for example, a 10$^\circ$ SNR of age 
10$^4$ years is seen from a distance of 200 pc and the expected flux
for an isolated SN is $\sim 1.7\cdot 10^{-11}{\rm cm}^{-2}{\rm
s}^{-1}$ for CR-ISM collisions, which is again less than the
minimum detectable flux of $3\cdot 10^{-11}{\rm cm}^{-2}{\rm
s}^{-1}$, although only by a factor of 1.8, which is within the
uncertainty of these estimates and the factor of 2.8 (~see \S2.3.3~)
to allow for heavier nuclei in CR and the ISM would be
sufficient to cover this difference.  

For the SNR similar to our 'Single Source' there is another
possibility to reduce the discrepancy. As discussed in
\ref{sec:emisA} the intensity of gamma rays is dependent on
the mass of the initiating particle; specifically for a `thin
target' (~the gas in the SNR~) and for a fixed rigidity spectrum it is 
proportional to {\cal Z} (~equation
\ref{eq:emisA}~). Thus,
{\em if} the majority of the particles were `heavy nuclei'---say
oxygen and iron (as required to explain the knee in the Single Source 
Model)---then the gamma ray emission would be enhanced.

To re-iterate, a nearby, isolated SNR, producing mainly heavy
nuclei at TeV energies and viewed against a low Galactic
background might just be detectable; an increase in CR energy,
too, would take it into the observable class. 

In what follows we examine the detected TeV gamma ray sources,
which seem to be important for our analysis.

\subsubsection{The detected TeV gamma ray sources}
\label{sec:detTeV}
All three shell type SNR detected to date in TeV gamma rays are
relatively distant sources and the sensitivity of detection is a maximum.
  
(i) CasA. As remarked in Section~\ref{sec:TeV-SNR}, CasA (~3C 461~)
has a measured flux
of $6\times10^{-13}\ {\rm cm}^{-2}{\rm s}^{-1}$ above
1~TeV. For a distance of 3.4 kpc, Figure 4 indicates an expected gamma
ray flux of $2.6\cdot 10^{-14}$cm$^{-2}$s$^{-1}$ for $n = 1$cm$^{-3}$,
and for our preferred density of $n = 0.1$cm$^{-3}$ the flux would
fall to $2.6\cdot 10^{-15}$cm$^{-2}$s$^{-1}$. An increased CR yield by
a factor 8 (~see \S8~) we would predict $\sim2\cdot
10^{-14}$cm$^{-2}$s$^{-1}$, a value too low by a factor of 30. Before 
dismissing $\pi^\circ$-decay as the source of the detected gamma rays
mention must be made of very recent work by Berezhko et
al. (~2003~). These authors adopt a mean density of 11 cm$^{-3}$ (~see
\S4.3~) and, with $0.6\cdot 10^{50}$erg going into CR derive an
expected flux equal to that observed. Our own model applied using
these parameters (~11 cm$^{-3}$ and $0.6\cdot 10^{50}$erg~) would give
a similar result - a factor 2 smaller. With the factor 2.8 from
\S2.3.3 there is near agreement.

The density is so high because the progenitor is assumed to have
injected 2M$_\odot$ into very small \\(~2 pc radius~) SNR. Whether this
is reasonable remains to be seen.

(ii) SN 1006. For the source, SN1006, the observed flux is 
$\sim 6\times10^{-12}\ {\rm cm}^{-2}{\rm s}^{-1}$, to be compared with 
expectation (~Figure 4~) of $\sim 9\cdot 10^{-14} 
\ {\rm cm}^{-2} {\rm s}^{-1}$ for $n=0.1\ {\rm cm}^{-3}$. At first
sight, explanation of the deficit by way of IC appears to be the
solution but we wish to point out the conclusions of Berezhko et
al. (~2001~) - denoted BKV. These authors maintain that the IC
contribution is only $\sim$30\% of the total expected intensity, the
bulk coming from $\pi^\circ$-decays. In passing, and this has
relevance to the discussion in \S1, they point out that the different
spatial distribution of low energy electrons which produce the radio
signal and the high energy electrons which can generate the TeV gamma
rays are very different (~the former being concentrated in a thin
shell, where the magnetic field is very strong~). Thus, the implied
high energy electron intensity is lower than would have been claimed
from the straightforward use of the radio data.

BKV use values for the SNR shock energy going into CR and gas density
higher than ours such that they are able to explain the whole signal
from SN 1006 in terms of $\pi^\circ$-decay. In fact, relaxing our
parameters to theirs would only increase our flux to $\sim 7\cdot
10^{-13}$ cm$^{-2}$s$^{-1}$, a factor 9 short. Part of the reason for
the difference in predictions of our model and that of BKV (~a factor
$\sim$3~) may come from differences in the spectral shape of the
proton component but the reason for the residual difference is
unknown (~application of the 2.8 factor from \S2.3.3 is not
appropriate because we are making a comparison~).

Notwithstanding the claim by BKV to be able to explain the TeV emission
by $\pi^\circ$-decay, we feel that IC may be the answer. The reason is
that the gamma ray emission comes from a restricted region of the
remnant only and it is unphysical to assume that the whole of the CR
injection is concentrated in this region.

(iii) SNR G347.3-0.5. As for the ROSAT source RX J1713.7-3946 
(~SNR G347.3-0.5~), possibly
associated with the EGRET GeV source 3EG J1714-3857, the measured
fluxes are $4.4 \cdot 10^{-7}{\rm cm}^{-2}{\rm s}^{-1}$ above 0.1 GeV 
and $1.0\cdot 10^{-11}{\rm cm}^{-2}{\rm s}^{-1}$ above 1 TeV. The
distance to this source is very uncertain, values quoted are within 1
- 6 kpc, and the age
is unknown. However, even with these uncertainties we can say that the
maximum expected fluxes are about $3\cdot 10^{-8}$ and $3 \cdot
10^{-12}{\rm cm}^{-2}{\rm s}^{-1}$ respectively for $n \approx 1{\rm
cm}^{-3}$, i.e. substantially less than observed, even for this `high'
density, but within the accessible region when a higher CR energy
content is allowed.

\section{Conclusions}

We have used our `standard' SNR model to predict gamma ray fluxes
above two limiting energies (0.1~GeV and 1~TeV) for remnants of
various ages and at different distances. Attention has been drawn
to the crucial question of the (ISM) gas into which the remnant
expands and it is pointed out that this can vary from one remnant
to another over the range $10^{-1}$ to $10^{-3}$
H-atoms~cm$^{-3}$ together with - in some cases - contributions from
molecular clouds or gas from the progenitor.

Comparison with observation shows our (~conventional~) predictions to be always
lower than observation (or the estimated upper limit to the
measured fluxes) and this is particularly so in the lower energy
region. In the TeV region the predictions are
nearer to observation but it is disturbing that where positive
observations have been made rather extreme assumptions must be
made. Loop I is, perhaps, an
exception insofar as there is approximate agreement,
without recourse to unusual features. It must be assumed that
processes other than proton (~and other nuclei~) - interstellar gas
interactions are mainly responsible for the measured gamma ray
fluxes, particularly at low energies.

The conclusion of relevance to cosmic ray origin is that, in view of
the predicted gamma ray fluxes never exceeding observation, there is,
as yet, no objection to particles to at least 10~TeV having been
accelerated by supernova remnants.

{\large{\bf Acknowledgments}}

The authors are grateful to The Royal Society and The University of
Durham for financial support. Two unknown referees are also thanked
for useful comments and suggestions.

\vfill 

\begin{table}[ht]
\begin{center}\begin{tabular}{ c c c c }
SNR  &  Age ($10^3$~y)  &  Distance (kpc)  &  Interacting? \\
\hline
IC443  &  3.0  &  0.7--2.0  &  Yes \\
MSH11-61A  &  2.2  &  2.2--518 (?)  &  No (?) \\
G 312.4-04  &  15.0  &  $\sim 5$  &  Yes (?) \\
W28  &  2.5 (?)  &  1.6--4.2  &  Yes \\
$\gamma$ Cygni  &  14  &  $\sim1.5$  &  Yes
\end{tabular}
\caption{\footnotesize Likely SNR--Gamma Ray Source Coincidences
(from Sturner et~al., 1996); $E_\gamma>0.1\ \GeV$.}
\end{center}\end{table}

\vspace{4ex}

\begin{table}[ht]
\begin{center}\begin{tabular}{ c c c c c c }
Name  &  SNR   &  distance   &  Ang. radius   &  Approx. radius  &  $I_\nu$(1\ GHz)  \\
  &  ($l,b$)  &  (kpc)  &  (deg.)  &  (pc)   &  (Jy) \\
\hline
---  &  G 65.3 +5.7  &  0.8  &  2.5  &  32  &  52 \\
Cygnus Loop  &  G 74.0 $-8.5$  &  0.44  &  2.0  &  14  &  210 \\
---  &  G 89.0 +4.7  &  0.8  &  1.0  &  13  &  220 \\
Monoceros  &  G 205.5 +0.5  &  0.8--1.6  &  1.8  &  30  &  160 \\
Vela  &  G 263.9 $-3.3$  &  0.25-0.5  &  2.0  &  1  & 1750
\end{tabular}
\caption{\footnotesize SNR within 1 kpc (distance, angular radius
and 1~GHz radio intensity from Green, 2000).}
\end{center}\end{table}

\section*{References}

Aharonian, F.A., Akhperjanian, A., Barrio, J. et al., 2001, (\em
Astron. Astrophys.}, {\bf 370}, 112 \\
Aharonian F.A., 2002, astro-ph/0209360, submitted to {\em Astron. and
Astrophys.} \\ 
Alpgard K. et al., 1982, {\em Phys. Lett. B}, {\bf 115} 71 \\
Arnal E.M. and Mirabel, F., 1991, {\em Astron. Astrophys.}, {\bf 250}, 171 \\
Axford W.I., 1981, {\em Proc.17th Int.Cosmic Ray Conf.(Paris)},{\bf
12} 155 \\
Berezhko E.G., Elshin V.K. and Ksenofontov L.T., 1996, {\em
J. Exp. Theor. Phys.}, {\bf 82} 1 \\
Berezhko E.G. 1999 {\em Private communication} \\
Berezhko, E.G., Ksenofontov, L.T. and V\"{o}lk, H.J., 2001, {\em
Proc. 27th Int. Cosm. Ray Conf., (Hamburg)}, {\bf 6}, 2489 \\
Berezhko, E.G., P\"{u}hlhofer, G. and V\"{o}lk, H.J., 2003,
astro-ph/0301205 \\
Berezinsky V.S. et al., 1984, Astrophysics of Cosmic Rays, Nauka,
Moscow \\
Bhadra, A., 2002, {\em J. Phys. G: Nucl. Part. Phys.}, {\bf 28}, 397
\\
Bhat, C.L. et al., 1985, {\em Nature}, {\bf 314}, 515 \\
Blitz, L., 1980, {\em Giant Molecular Clouds in the Galaxy},
Eds. P.M.Solomon and M.G.Edmunds, Pergamon Press, 1 \\
Bykov, A.M. et al., 2000, {\em Astrophys. J.}, {\bf 538}, 203 \\
Chevalier, R.A., 1999, {\em Astrophys. J.}, {\bf 511}, 798 \\ 
Dodds, D., Wolfendale, A.W. and Wdowczyk, J., 1976 {\em
Mon. Not. Roy. Astron. Soc.}, {\bf 176}, 345 \\
Drury, L.O'C., Aharonian F.A. and V\"{o}lk, 1994, {\em
Astron. Astrophys} {\bf 287} 959 \\
Dyson, J.E. and Williams, D.A., 1980, {\em Physics of the Interstellar
Medium}, Manchester Univ. Press \\
Enomoto R. et al., 2002, {\em Nature}, {\bf 416}, 823 \\
Erlykin, A.D. and Wolfendale, A.W., 1997, {\em J. Phys. G:
Nucl. Part. Phys. } {\bf 23} 979 \\
Erlykin, A.D. and Wolfendale, A.W., 2001, {\em J. Phys. G:
Nucl. Part. Phys. } {\bf 27} 941 \\
Erlykin A.D., Lagutin A.A. and Wolfendale A.W. 2003 {\em
Astropart. Phys} (in press), astro-ph/0209506 \\
Erlykin A.D. and Wolfendale A.W., 2002a, {\em J. Phys. G:
Nucl. Part. Phys.} {\bf 28} 2329 \\
Erlykin A.D. and Wolfendale A.W., 2002b, {\em J. Phys. G:
Nucl. Part. Phys.} {\bf 28} 359 \\
Ferriere, K.M., 2001, astro-ph/0106359 \\
Frisch P., 1997, astro-ph/9705231 \\
Green, D.A., 2000, {\em A Catalogue of Galactic SNR}, Dept. of
Physics, Univ. of Cambridge, UK \\
Hara, S. et al., 2001, {\em Proc. 27th Int. Cosm. Ray Conf. (Hamburg)},
{\bf 6}, 2455 \\
Hartman, R.C. et al., 1999, {\em Astron. Astrophys. Suppl. Ser.}, {\bf
123}, 179 \\
Heiles, C., 1987, {\em Astrophys. J.}, {\bf 315}, 555 \\
Heiles, C., 1990, {\em Astrophys. J.}, {\bf 354}, 483 \\
He\ss, M. et al., 1997, {\em Proc. 25th
Int. Cosm. Ray. Conf. (Durban)},
{\bf 3} 229 \\
Houston, B.P. and Wolfendale, A.W., 1984, {\em J. Phys. G:
Nucl. Part. Phys.}, {\bf 10}, 1587 \\
Ito K., 1988, in {\em Cosmic Ray Astrophysics}, ed. by Oda M.,
Nishimure J. and Sakurai K., Terra Sci.Publ.Com./Tokyo \\
Lagutin A.A. et al., 2001a, {\em Nucl. Phys. B (Proc.Suppl.)} {\bf
97} 267 \\
Lagutin A.A. et al., 2001b, {\em Proc. 27th Int. Cosmic Ray
Conf.(Hamburg)}, {\bf 5} 1900 \\
Lampeitl H. et al., 2001, {\em Proc. 27th Int. Cosm. Ray Conf.,
(Hamburg)}, {\bf 6}, 2348 \\
Lebrun, F. and Paul, J., 1985, {\em Proc. 19th Int. Cosm. Ray Conf.,
(La Jolla)}, {\bf 1}, 309 \\
Li, T.P. and Wolfendale A.W. 1981, {\em Astron. Astrophys. Lett.},
{\bf 100}, L126 \\
Longair M.S. 1992, {\em High Energy Astrophysics}, Cambridge Univ. Press,
second. ed., vol.1 \\
Lozinskaya, T.A., 1991, {\em Supernovae and Stellar Winds in the
Interstellar Medium}, Amer. Inst. Phys. \\
Moffett, D.A., Gross, W.M. and Reynolds, S.P., 1993, {\em
Astrophys. J.}, {\bf 106}, 1566 \\
Osborne, J.L., Wolfendale, A.W. and Zhang, L., 1995, 
{\em J. Phys. G: Nucl. Part. Phys.}, {\bf 21}, 429 \\
Pare E. et al. 1990, {\em Phys. Lett. B} {\bf 242} 531 \\
Parizot, E., Paul, J. and Bykov, A., 2001, {\em Proc. 27th
Int. Cosm. Ray Conf. (Hamburg)}, {\bf 6}, 2070 \\ 
Plaga R. 2001, astro-ph/0111555, submitted to New Astronomy Review \\
Plucinsky P.P. et al., 1996, {\em Astrophys. J.}, {\bf 463}, 224 \\
Pohl M. 2001, {\em Proc. 27th Int.Cosmic Ray Conf. (Hamburg)},
Invited, Rapporteur and Highlight Papers, 147 \\
Pollock, A.M.T., 1985, {\em Astron. and Astrophys.}, {\bf 150}, 339 \\
Prosch C. et al., 1996, {\em Astron. and Astrophys.}, {\bf 314}, 275 \\
Ramana Murthy, P.V. and Wolfendale, A.W., 1993, {\em Gamma Ray
Astronomy}, Cambridge Univ. Press \\
Rogers, M.J. and Wolfendale A.W., 1987, {\em Proc. 20th Int. Cosm. Ray
Conf., (Moscow)}, {\bf 1}, 81 \\
Rossi B., 1952, {\em High Energy Particles}, Prentice Hall, Englewood
Cliffs, NJ \\
Stecker, F.W., 1971, {\em Cosmic Gamma Rays}, Mono Book Corp.,
Baltimore MD \\
Strong, A.W. and Mattox, J.R., 1996, {\em Astron. Astrophys.}, {\bf
308}, L21 \\
Sturner, S.J., and Dermer, C.D. 1995, {\em Astron. Astrophys.}, {\bf
293}, L17 \\
Sturner, S.J., Dermer, C.D. and Mattox, J.R., 1996 {\em
Astron. Astrophys. Suppl. Ser.}, {\bf 120}, 445 \\
Torres, D.F., Romero, G.F., Dame, T.M. et al., 2002,
astro-ph/0209565, submitted to Physics Reports \\
Torres, D.F. et al., 2003, astro-ph/0301424 \\
V\"{o}lk H., 2002, astro-ph/0210297 \\
Wallace, B.J., Landecker, T.L. and Taylor, A.R., 1994, {\em
Astron. Astrophys}, {\bf 286}, 565 \\
Wiebel-Sooth B. and Biermann P.L. 1999, {\em Astronomy and
Astrophysics - Interstellar Matter, Galaxy, Universe}, {\bf
3C}, 37, publ. in Landolt-B\"{o}rnstein, Group VI, {\em ISBN
0942-8011/3-540-56081-5}, Springer Verlag, Berlin/Heidelberg \\
Wolfendale, A.W. and Zhang, L., 1994, {\em J. Phys. G:
Nucl. Part. Phys.}, {\bf 20}, 935 
\end{document}